\documentclass[10pt,journal]{IEEEtran}

\usepackage{graphicx}
\usepackage{algorithm}
\usepackage{algorithmicx}
\usepackage{algpseudocode}
\usepackage{tabularx}
\usepackage{graphicx}
\usepackage{caption}
\usepackage{subcaption}

\usepackage{makecell}
\usepackage{amsmath}

\usepackage{enumitem}
\usepackage[hyphens]{url}
\ifCLASSINFOpdf
\else
\fi
\begin{document}
%
\title{GenDRAM:Hardware-Software Co-Design of General Platform in DRAM}

\author{Tsung-Han~Lu,Weihong~Xu,~\IEEEmembership{Member,~IEEE},and~Tajana~Rosing,~\IEEEmembership{Fellow,~IEEE}
\vspace{-0.9cm}

\thanks{T.H Lu, and T. Rosing are with the Department of Computer Science and Engineering, University of California San Diego, La Jolla, CA, 92093 USA (email: \{tsl012, tajana\}@ucsd.edu).}
\thanks{W. Xu with the Department of Electrical and Computer Engineering, Swiss Federal Institute of Technology in Lausanne (EPFL) (email: \{weihong.xu\}@epfl.ch).}
}

\maketitle

\begin{abstract}
Dynamic programming (DP) algorithms, such as All-Pairs Shortest Path (APSP) and genomic sequence alignment, are fundamental to many scientific domains but are severely bottlenecked by data movement on conventional architectures. While Processing-in-Memory (PIM) offers a promising solution, existing accelerators often address only a fraction of the workflow, creating new system-level bottlenecks in host-accelerator communication and off-chip data streaming. In this work, we propose GenDRAM, a massively parallel PIM accelerator that overcomes these limitations. GenDRAM leverages the immense capacity and internal bandwidth of monolithic 3D DRAM(M3D DRAM) to integrate entire data-intensive pipelines, such as the full genomics workflow from seeding to alignment, onto a single heterogeneous chip. At its core is a novel architecture featuring specialized Search PUs for memory-intensive tasks and universal, multiplier-less Compute PUs for diverse DP calculations. This is enabled by a 3D-aware data mapping strategy that exploits the tiered latency of M3D DRAM for performance optimization. Through comprehensive simulation, we demonstrate that GenDRAM achieves a transformative performance leap, outperforming state-of-the-art GPU systems by over 68$\times$ on APSP and over 22$\times$ on the end-to-end genomics pipeline.
\end{abstract}


%
\IEEEpeerreviewmaketitle

\section{Introduction}
All-Pairs Shortest Path (APSP) stands as a fundamental algorithm in domains such as network routing, logistics, sequence alignment and social network analysis ~\cite{APSPApradhan2013finding, APSPArout2024systematic, APSPAgutman2004reach}. The canonical Floyd-Warshall (FW) algorithm ~\cite{floyd1962algorithm}, however, presents a significant computational challenge with its $O(N^{3})$ time complexity and the massive data movement required to repeatedly update its dense distance matrix. While GPU-based implementations have offered performance improvements over CPUs ~\cite{sao2021scalableGPU, garland2008parallelCUDA, djidjev2014efficient}, they remain fundamentally limited by the data-movement bottleneck between processing cores and off-chip memory, which dominates both execution time and energy consumption. The tiled structure and inherent parallelism of the blocked FW algorithm make it an ideal candidate for Processing-in-Memory (PIM) architectures ~\cite{zhou2022transpim, ahn2015scalablePIM,chi2016prime}, which aim to resolve the memory wall by embedding computation directly within the memory hierarchy.

Concurrently, the advancement of high-throughput sequencing (HTS) technologies has revolutionized fields ranging from precision medicine to epidemiology ~\cite{khaleghi2022salient}. Modern sequencers can now generate terabytes of genomic data in mere hours, creating a data deluge that dramatically outpaces the scaling predicted by Moore's Law ~\cite{khaleghi2022salient}. Within this complex bioinformatics pipeline, a critical and often dominant bottleneck is sequence alignment, a computationally demanding task that can consume 60-80\% of the total runtime ~\cite{xu2023rapidx}. This escalating computational challenge, much like that in graph analytics, necessitates a fundamental paradigm shift in hardware accelerator design. Crucially, both APSP and alignment are foundationally rooted in Dynamic Programming (DP), sharing a computational core that relies heavily on iterative neighbor updates and minimal arithmetic units (adders and comparators). This shared DP nature serves as the primary motivation for developing a single, unified platform capable of accelerating both graph and genomic analysis, maximizing hardware utilization.

To address these diverse, memory-intensive workloads, numerous hardware accelerators have been proposed on ASIC, GPU, and FPGA platforms ~\cite{sao2021scalableGPU,chirila2022heterogeneousFPGA}. More recently, PIM has emerged as a compelling solution to mitigate the profound data movement costs ~\cite{dai2018graphh, song2018graphr,kim2023samsung}. State-of-the-art PIM accelerators like RAPIDx have demonstrated significant throughput improvements for sequence alignment by co-designing novel algorithms with ReRAM-based hardware ~\cite{xu2023rapidx}. Similarly, FPGA-based designs like SALIENT have achieved high performance by creating specialized, pipelined hardware for different stages of the alignment workflow ~\cite{khaleghi2022salient}. However, these advanced solutions still face fundamental limitations rooted in system-level bottlenecks. Due to on-chip memory capacity constraints, many designs cannot store the entire multi-gigabyte reference genome, forcing them to frequently stream reference data from a host system and re-introducing a significant data movement bottleneck. Furthermore, to simplify the accelerator design, critical pre-processing steps like seeding are often offloaded to the host CPU. This division of labor breaks the analysis pipeline and creates a new, often overlooked, system-level bottleneck in the communication between the host and the accelerator ~\cite{pan2025stratum}. 

The limitations of prior PIM designs underscore the necessity for a truly monolithic, end-to-end solution. Existing approaches often fail due to three key systemic flaws: (1) Incomplete pipelining: offloading memory-intensive pre-processing stages, such as the seeding phase in genomics, to the host CPU re-introduces a critical, system-level communication bottleneck. (2) Capacity constraints: many accelerators lack the on-chip memory capacity to store massive structures like the entire multi-gigabyte reference genome, necessitating inefficient off-chip data streaming. (3) Lack of universality: prior PIM solutions are often highly specialized ~\cite{xu2023rapidx, dai2018graphh, song2018graphr}, addressing only a single DP variant, thus failing to maximize hardware utilization across the increasingly diverse landscape of data-intensive workloads. 


In this work, we propose GenDRAM, a massively parallel 3D Processing-in-Memory (PIM) accelerator designed to overcome system-level limitations across multiple domains. It unifies structurally disparate yet mathematically isomorphic tile-based dynamic programming workloads, ranging from the global systolic broadcasting of APSP to the local wavefront propagation of genomic alignment. To achieve this, GenDRAM has a heterogeneous architecture capable of accelerating a broad spectrum of memory-bound scientific computing bottlenecks rooted in the generalized grid update problem.  The architecture's key contribution is a novel heterogeneous on-chip pipeline featuring two distinct types of Processing Elements (PEs) to enable true pipeline parallelism, supported by a sophisticated 3D-aware data mapping strategy that exploits the tiered latency of M3D DRAM. Furthermore, we design a multiplier-less PE capable of efficiently executing the diverse operations required for both banded DP alignment and the FW APSP algorithm. Thus, leveraging the immense capacity and internal bandwidth of M3D DRAM, GenDRAM integrates the entire bioinformatics pipeline---from memory-intensive seeding to compute-intensive alignment---onto a single heterogeneous chip, thereby eliminating critical host-accelerator communication bottlenecks by co-locating specialized compute engines with large-scale data structures. Our comprehensive evaluation demonstrates that GenDRAM outperforms state-of-the-art GPU systems, achieving a \textbf{67$\times$ speedup on APSP} and a \textbf{22$\times$ speedup on the end-to-end genomics pipeline} compared to the NVIDIA A100, validating the efficacy of our proposed unified PIM architecture.

\section{Background}
\label{sec:background}

\subsection{APSP and Genomic Sequence Alignment}
\subsubsection {APSP and FW algorithm }
The APSP problem is a foundational algorithm in graph theory ~\cite{yang2023fast}, critical for diverse applications ranging from network routing to bioinformatics ~\cite{APSPApradhan2013finding, APSPArout2024systematic, APSPAgutman2004reach}. The canonical solution for dense graphs is the FW algorithm ~\cite{floyd1962algorithm}, which operates with a time complexity of $O(N^3)$. Its core mechanism involves iteratively relaxing edges through intermediate vertices $k$, updating distances via the equation $d(i, j) = \min(d(i, j), d(i, k) + d(k, j))$. While conceptually simple, the standard FW algorithm exhibits poor cache locality and inefficient data reuse on large matrices, creating significant performance bottlenecks in conventional CPU architectures.

\begin{figure}[htbp]
    \centering
    \includegraphics[width=\columnwidth]{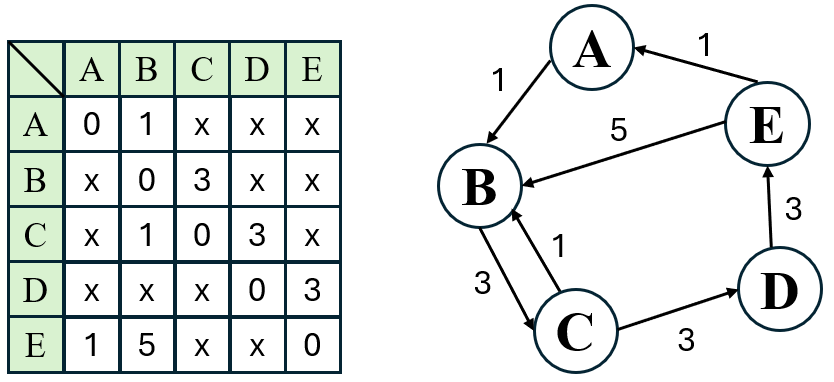} 
    \caption{Example of a weighted directed graph (right) and its corresponding adjacency matrix (left), serving as the input for the APSP problem}
    \label{fig:APSP_example}
\end{figure}

\begin{figure}[htbp]
    \centering
    \includegraphics[width=\columnwidth]{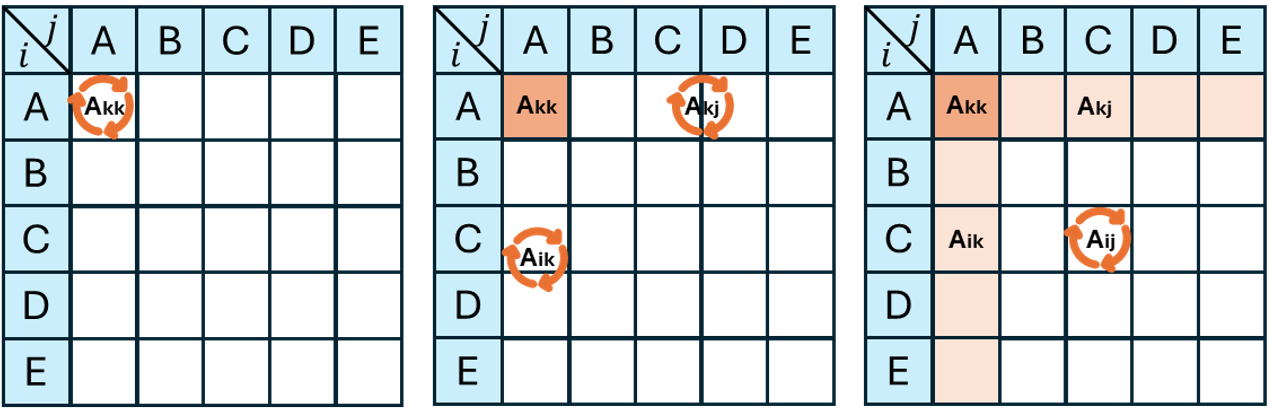} 
    \caption{Three phases of the Blocked FW algorithm for updating a distance matrix. Orange arrows indicate data dependencies or computation within a block or between blocks and pivot elements}
    \label{fig:blocked_fw}
\end{figure}

To mitigate these inefficiencies, the blocked FW algorithm ~\cite{prihozhy2023speedupblockedFW, likhoded2019generalizedblockedFW} partitions the $N \times N$ distance matrix into $B \times B$ tiles sized to fit within on-chip memory. As illustrated in Fig. \ref{fig:blocked_fw}, execution proceeds in $N/B$ super-steps, each consisting of three distinct phases: (1) \textbf{Self-update}, where the pivot block $A_{kk}$ is updated sequentially; (2) \textbf{Row/column update}, where blocks in the pivot row $A_{kj}$ and column $A_{ik}$ are updated using $A_{kk}$; and (3) \textbf{Internal update}, where all remaining blocks $A_{ij}$ are updated in parallel using the respective pivot row and column blocks.

This tiled approach optimizes data access patterns by maximizing locality within each block processing step, as outlined in Algorithm \ref{alg:blocked_fw}. By completing updates on local tiles before advancing, the blocked FW algorithm significantly reduces off-chip data movement~\cite{prihozhy2023speedupblockedFW}. This characteristic renders it particularly suitable for PIM architectures, where Compute PEs can efficiently process data blocks streamed directly from underlying DRAM layers, thereby overcoming the memory wall.

\begin{algorithm}[h!]
\footnotesize
\caption{Blocked Floyd-Warshall Algorithm}
\label{alg:blocked_fw}
\begin{algorithmic}[1]
\State \textbf{Input:} Weighted graph matrix $D$ of size $N \times N$, block size $B$.
\State \textbf{Output:} APSP distance matrix $D$.
\State $num\_blocks \gets N/B$
\For{$k \gets 0$ to $num\_blocks-1$}
    \Comment{Phase 1: Self-Update of the pivot block}
    \State $FW\_on\_block(D[k][k])$

    \Comment{Phase 2: Update pivot row and column}
    \For{$i \gets 0$ to $num\_blocks-1$}
        \If{$i \neq k$}
            \State $Block\_Update(D[i][k], D[i][k], D[k][k])$
        \EndIf
    \EndFor
    \For{$j \gets 0$ to $num\_blocks-1$}
        \If{$j \neq k$}
            \State $Block\_Update(D[k][j], D[k][k], D[k][j])$
        \EndIf
    \EndFor

    \Comment{Phase 3: Update internal blocks}
    \For{$i \gets 0$ to $num\_blocks-1$}
        \For{$j \gets 0$ to $num\_blocks-1$}
            \If{$i \neq k$ and $j \neq k$}
                \State $Block\_Update(D[i][j], D[i][k], D[k][j])$
            \EndIf
        \EndFor
    \EndFor
\EndFor
\State \textbf{return} $D$
\end{algorithmic}
\end{algorithm}

\subsubsection { Genomic Alignment}

\begin{figure}[htbp]
    \centering
    \includegraphics[width=\columnwidth]{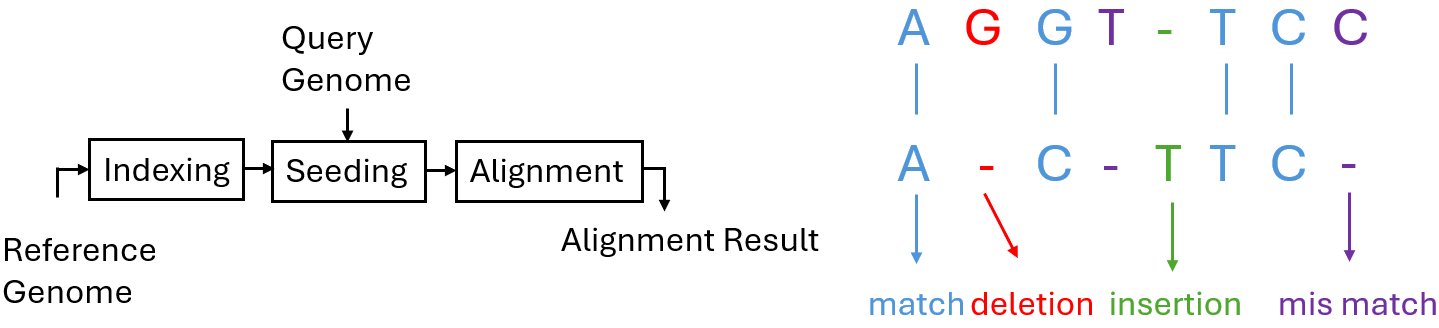} 
    \caption{Overview of the genomic sequence alignment pipeline (left) and illustration of basic operations in sequence alignment (right): match, deletion, insertion, and mismatch. These operations are associated with different scores or penalties in dynamic programming algorithms}
    \label{fig:genomic_alignment_overview}
\end{figure}

Genomic sequence alignment is a computationally intensive cornerstone of bioinformatics, requiring the mapping of millions of short DNA reads to a reference genome. This process is often bottlenecked by memory and compute limitations. Modern aligners mitigate this complexity through a multi-stage pipeline, as depicted in Fig. \ref{fig:genomic_alignment_overview}. The pipeline typically begins with indexing to build rapid lookup structures and seeding, a memory-intensive phase that identifies potential alignment locations by matching subsequences against the index. Following a filtering step to prune candidates, the most compute-intensive phase, alignment, employs DP algorithms like Smith-Waterman to compute optimal alignment scores~\cite{xu2023rapidx}.

Throughout this work, we define the end-to-end genomics workflow as the online processing stages: \textbf{seeding} and \textbf{alignment} (including filtering and traceback). The indexing stage, which builds the hash tables (PTR/CAL), is a one-time offline pre-processing step performed on the host and is therefore excluded from runtime measurements. Similarly, final I/O formatting is excluded as it is not on the critical path of the accelerator. This definition aligns with standard benchmarking practices where the read mapping throughput is the primary metric of interest~\cite{khaleghi2022salient, xu2023rapidx}.

\begin{figure}[htbp]
    \centering
    \includegraphics[width=1 \columnwidth]{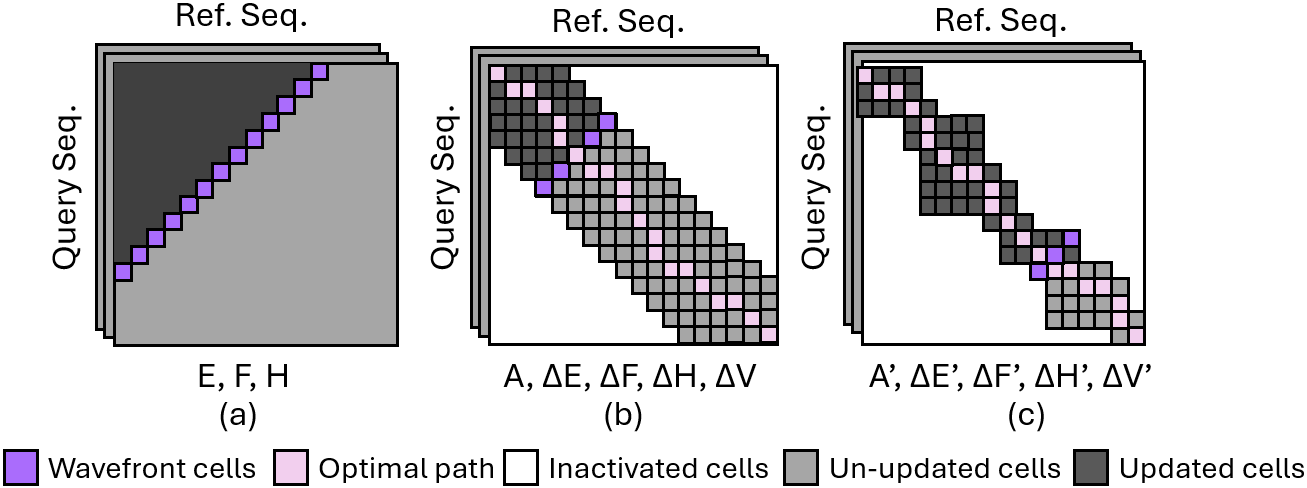} 
    \caption{Illustration of three variants of DP alignment algorithms. (a) Original full DP. (b) Banded difference-based DP. (c) Adaptive banded parallelized DP~\cite{xu2023rapidx}}
    \label{fig:alignment_band_rapidx}
\end{figure}

A critical optimization for DP-based alignment is \textbf{banded dynamic programming}, which reduces the computational complexity from quadratic $O(L_{\text{read}} \times L_{\text{reference}})$ to linear $O(L_{\text{read}} \times \mathit{BandWidth})$ by restricting computation to a narrow diagonal band~\cite{xu2023rapidx}. As illustrated in Fig. \ref{fig:alignment_band_rapidx}, advanced accelerators like RAPIDx~\cite{xu2023rapidx} further refine this approach using difference-based and adaptive techniques. Unlike the prohibitively large original full DP, the banded difference-based DP restricts computation to a fixed bandwidth (e.g., $B=6$), computing score differences rather than absolute values to simplify hardware. The adaptive banded parallelized DP further optimizes this by dynamically adjusting a reduced bandwidth (e.g., $B=3$), minimizing memory access and computation for highly similar sequences. This approach is fundamental for high-throughput hardware acceleration. However, while RAPIDx provides an efficient algorithmic framework, its hardware implementation relies on ReRAM, which has much slower writes, low reliability and challenges with scaling. In contrast, GenDRAM uniquely adapts this adaptive banded algorithm to M3D DRAM. By coupling this algorithmic efficiency with the massive capacity with fast latency and internal bandwidth of M3D DRAM, GenDRAM enables a truly monolithic, end-to-end accelerator design.

This optimization strategy aligns with the \textit{Blocking} or \textit{Tiling} strategy used in APSP. Although sequence alignment (local wavefront) and Floyd-Warshall (global broadcast) exhibit different dataflow patterns, both are canonical DP applications relying on a shared, tile-based recursive update model (\texttt{max} vs. \texttt{min-plus}). This shared paradigm is the key enabler for our unified PIM accelerator.

The selection of M3D DRAM as the foundational technology for GenDRAM is driven by the unique requirements of these workloads. While alternative PIM technologies like ReRAM offer computation density, they lack the capacity for genomic workloads~\cite{xu2023rapidx}. Conversely, conventional M3D DRAMs like HBM suffer from internal bandwidth bottlenecks due to TSV interconnects~\cite{pan2025stratum}. M3D DRAM uniquely bridges this gap, providing DRAM-scale capacity and massive internal bandwidth via hybrid bonding~\cite{pan2025stratum}, essential for sustaining the massive parallelism in our GenDRAM design.

\subsection{Unified View of Dynamic Programming Workloads}
\label{sec:unified_view}

While APSP and genomic sequence alignment originate from distinct scientific domains, they share a fundamental algorithmic structure rooted in DP on dense matrices. We abstract both problems as a \textit{generalized grid update} problem, characterized by two key isomorphisms:

\subsubsection{Computational isomorphism}
\begin{figure}[htbp]
    \centering
    \includegraphics[width=1\columnwidth]{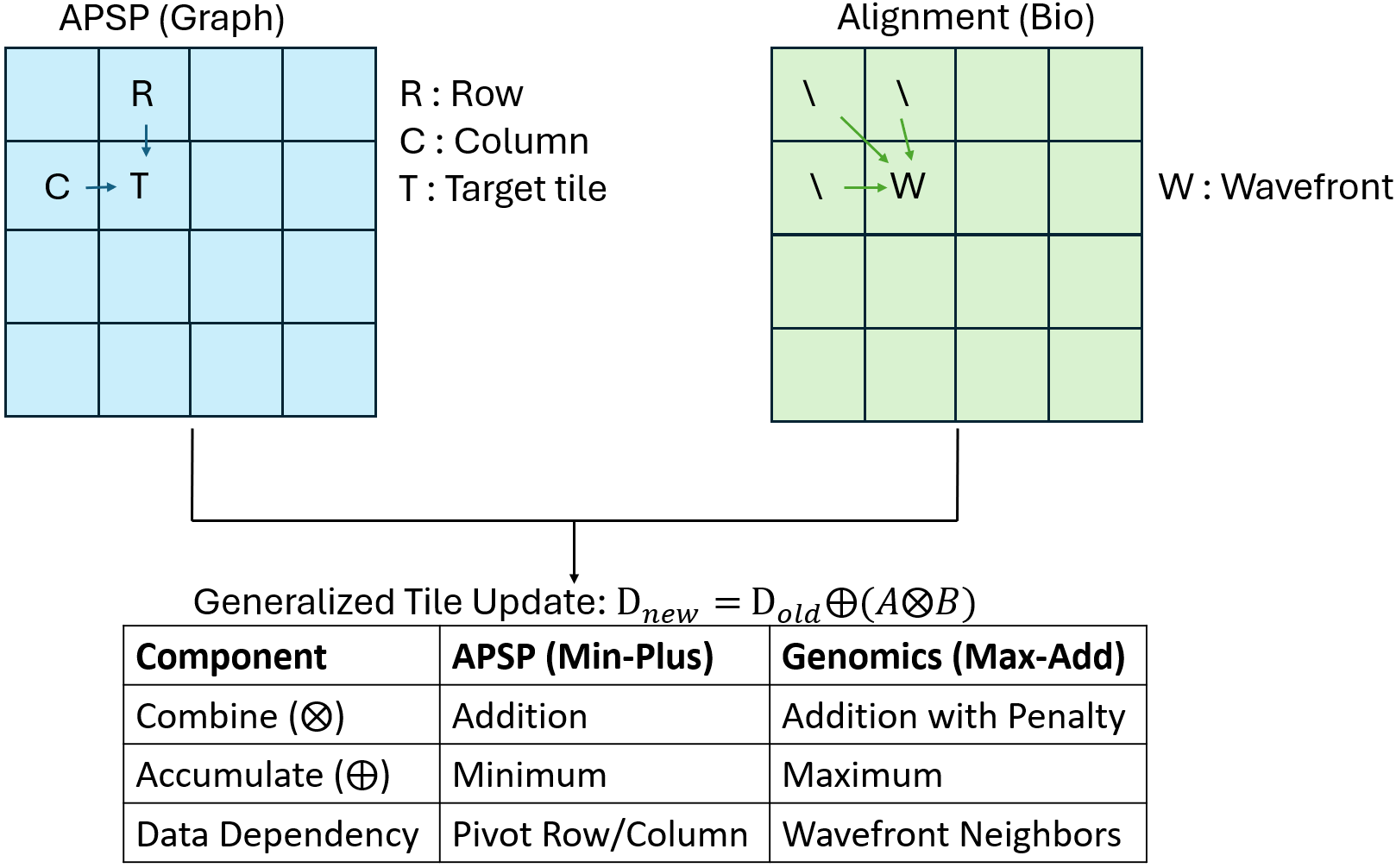} 
    \caption{The GenDRAM Unified Computing Abstraction. Despite differing data access patterns (Global Broadcast in APSP vs. Local Wavefront in Genomics), both workloads map to a unified Generalized Grid Update problem based on semi-ring algebra}
    \label{fig:unified}
\end{figure}

Both algorithms involve iteratively updating a target cell $D[i,j]$ based on data from neighbor cells or pivot blocks. This operation can be generalized as a semi-ring operation over a matrix:
\begin{equation}
    D[i,j] \leftarrow D[i,j] \oplus (D[i,k] \otimes D[k,j])
\end{equation}
where $\oplus$ represents the accumulation operator and $\otimes$ represents the combination operator as shown in Fig.~\ref{fig:unified}.
\begin{itemize}
    \item \textbf{For APSP (Floyd-Warshall):} The problem is defined over the $(\min, +)$ semi-ring. The system minimizes the path cost ($\oplus \rightarrow \min$) by summing edge weights ($\otimes \rightarrow +$).
    \item \textbf{For Sequence alignment (Smith-Waterman):} The problem is defined over the $(\max, +)$ semi-ring. The system maximizes the alignment score ($\oplus \rightarrow \max$) by adding match/gap penalties ($\otimes \rightarrow +$).
\end{itemize}

\noindent \textbf{The Data movement bottleneck:} Despite the algorithmic simplicity, these workloads are severely throttled by the memory wall on conventional architectures. 
For genomic alignment, the Operational Intensity (OI) is inherently low due to frequent backtracking and table lookups, with profiling data showing that memory stalls account for 60\%--80\% of the total runtime in standard software aligners~\cite{xu2023rapidx}. 
Similarly for APSP, while the complexity is cubic, the energy cost is dominated by data movement. Fetching operands from off-chip DRAM consumes approximately 200$\times$ more energy (approx. 640 pJ per 64-bit access) compared to the simple integer arithmetic operations (approx. 3 pJ per add) required for the update~\cite{pan2025stratum}. 
This quantitative disparity---where data movement cost dwarfs computation cost---motivates the need for a PIM architecture that minimizes distance-to-data.
This quantitative disparity---where data movement cost dwarfs computation cost---motivates the need for a PIM architecture that minimizes distance-to-data.

\subsubsection{Structural commonality}
To mitigate the memory wall, both algorithms utilize \textit{blocked (tiled) execution}. Whether it is the wavefront propagation in alignment or the pivot-row-column update in blocked FW, both workloads require fetching a $B \times B$ tile, performing arithmetic on local registers, and broadcasting dependencies to adjacent tiles. 

This structural and computational isomorphism implies that a single hardware substrate---equipped with reconfigurable $(\oplus, \otimes)$ Arithmetic units and a mesh/ring interconnect for tile broadcasting---can universally accelerate both workloads without the need for dedicated, disjoint ASIC logic, which is the motivation for this work.

\subsection{Monolithic 3D-Stackable DRAM}
\begin{figure}[htbp]
    \centering
    \includegraphics[width=1\columnwidth]{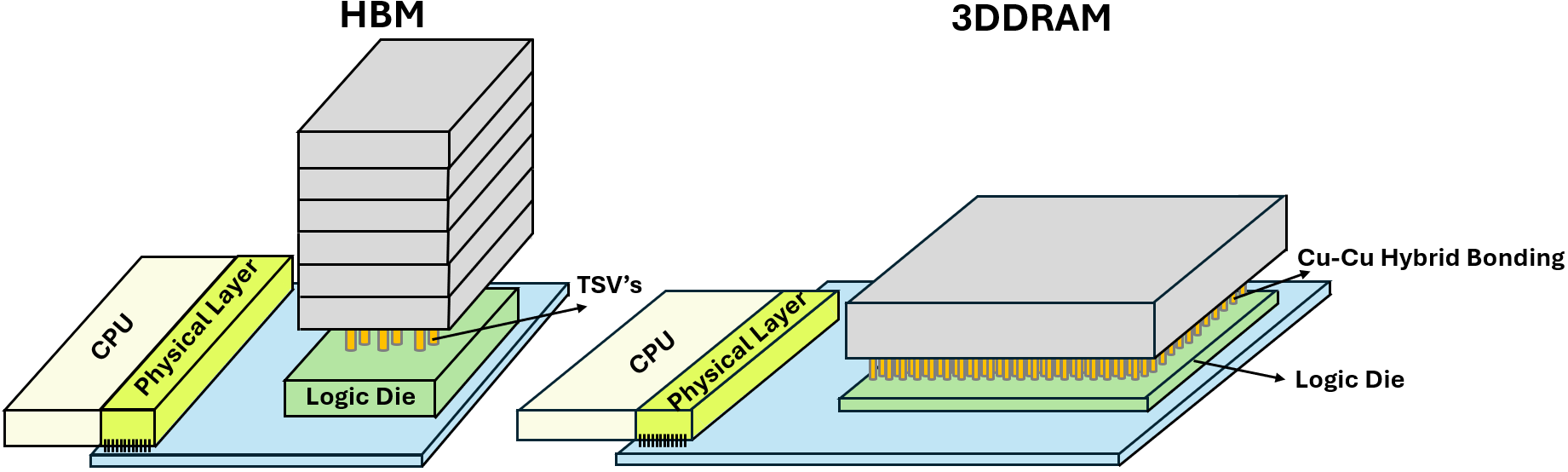} 
    \caption{Architectural comparison between conventional HBM (left)~\cite{jun2017hbm} using TSVs and M3D DRAM (right)~\cite{hsu20253DDRAM} using Cu-Cu Hybrid Bonding for vertical interconnection}
    \label{fig:hbm_vs_mono3d_comparison}
\end{figure}

Our architecture is founded on M3D DRAM ~\cite{hsu20253DDRAM}, which offers fundamental advantages over conventional High-Bandwidth Memory (HBM) for PIM systems. While HBM relies on coarse-pitch (\textasciitilde$10~\mu m$) Through-Silicon Vias (TSVs) that create a critical bottleneck by limiting internal bandwidth and density (Fig. \ref{fig:hbm_vs_mono3d_comparison}), M3D DRAM utilizes order of magnitude finer pitch (\textasciitilde$1~\mu m$) Cu-Cu hybrid bonding interface. This architectural distinction enables an order-of-magnitude denser vertical connectivity, unlocking tens of TB/s of internal bandwidth---achieved via massive parallelism across numerous banks and channels ---and supporting a higher power budget for co-located logic due to improved thermal performance. We selected M3D DRAM as its massive internal bandwidth, superior integration density, higher capacity, and exploitable characteristics are the key enablers for our high-throughput PIM pipeline.

\begin{figure}[htbp]
    \centering
    \includegraphics[width=1.1\columnwidth]{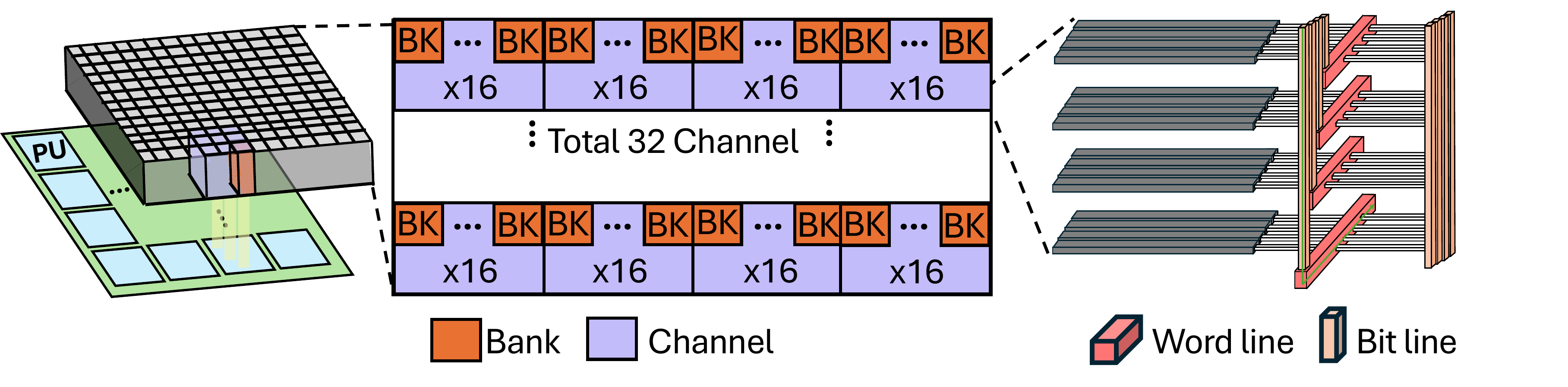} 
    \caption{Illustration of the M3D DRAM architecture. The left side shows multiple Processing Units (PUs) on the Logic Die and the hierarchical organization of DRAM into Banks and Channels. The magnified view on the right details the Wordline (WL) routing, demonstrating its staircase structure across multiple DRAM layers}
    \label{fig:mono3d_wordlines}
\end{figure}

A key physical characteristic of aggressively scaled M3D DRAM is the emergence of heterogeneous access latencies across its vertical layers. As illustrated in Fig. \ref{fig:mono3d_wordlines}, wordlines (WLs) are routed along a staircase structure, resulting in a linearly increasing RC delay for layers farther from the logic die . This creates a tiered latency profile where bottom layers exhibit the fastest access times. GenDRAM is designed to exploit this: rather than a limitation, this inherent heterogeneity presents a unique optimization opportunity. We employ a 3D-aware data mapping strategy to strategically place performance-critical hot data into the fast, low-latency tiers, thereby maximizing average memory access performance.

\subsection {Challenges on M3D DRAM for Dynamic Programming}
While M3D DRAM offers unprecedented capacity and internal bandwidth, its unique physical characteristics present significant challenges for accelerating dynamic programming algorithms like APSP and sequence alignment. 

\begin{enumerate} [leftmargin=*]
    \item \textbf{Different algorithmic access patterns:}
    The first challenge originates from the DP algorithms themselves, which feature complex, data-dependent access patterns. This is far removed from simple linear data scans. Instead, the hardware must support two fundamentally different and dynamically shifting dataflow modes: the wavefront propagation required by genomic alignment, and the pivot block access with broadcasting essential to blocked FW. This fundamental algorithmic irregularity makes it exceptionally difficult to design a single, optimal static data layout for both, thereby placing high demands on the hardware's real-time dataflow control.

    \item \textbf{3D mapping and tiered latency:}
    The physical structure of M3D DRAM presents a second major challenge, combining the complexities of tiered latency and data mapping~\cite{pan2025stratum}. First, due to the wordline staircase structure, DRAM layers farther from the logic die exhibit 
    significantly longer access latencies. This forces a vertical optimization, where critical data (like a pivot block or active wavefront) must be strategically placed in low-latency fast tiers.The physical structure of M3D DRAM presents a second major challenge: balancing the trade-off between tiered access latency and memory parallelism~\cite{pan2025stratum}. Due to the staircase wordline structure, DRAM layers positioned farther from the logic die exhibit significantly longer access latencies. This characteristic necessitates a vertical optimization, where latency-critical data must be strategically mapped to the faster, bottom-most tiers. However, this vertical constraint stands in tension with the goal of horizontal optimization, which seeks to maximize bandwidth by interleaving data blocks across disparate channels and banks. Simultaneously satisfying both objectives is non-trivial; aggressively concentrating hot data into a few fast tiers can restrict the available bank-level parallelism, potentially creating access congestion within those specific channels. We must simultaneously maximize data locality (the vertical optimization)  and maximize memory parallelism (a horizontal optimization, by interleaving data blocks across different channels or banks). Achieving both goals simultaneously in 3D space is non-trivial, as concentrating all hot data in fast tiers might create horizontal access congestion, and vice versa.

    \item \textbf{Compute PU design challenge:}
    Finally, designing a single PU to efficiently accelerate both algorithms presents a critical conflict in functional structure and data precision. The PU must be reconfigurable to support two distinct datapath structures: APSP's \texttt{min-plus} (\texttt{min(A, B+C)}) and genomic alignment's \texttt{max-logic} (\texttt{max(A, B, C+D)}). The more severe challenge lies in the starkly divergent data precision: APSP's path accumulation \textit{requires} a wide datapath (e.g., 32-bit) to prevent overflow, whereas the adaptive banded DP algorithm for sequence alignment is optimized for an ultra-low 5-bit precision~\cite{xu2023rapidx}. This creates a critical design trade-off: a PU optimized for 5-bit is functionally incorrect for 32-bit tasks. Conversely, a 32-bit PU, while capable of handling both, is massively inefficient for the 5-bit workload, wasting over 84\% of its resources. The architectural challenge, therefore, is to select a unified PU design that correctly balances these conflicting demands of functional correctness, universality, and computational efficiency.
\end{enumerate}

 
\section{GenDRAM Hardware Architecture}
\subsection{GenDRAM Overview}
\begin{figure}[htbp]
    \centering
    \includegraphics[width=1.0\columnwidth]{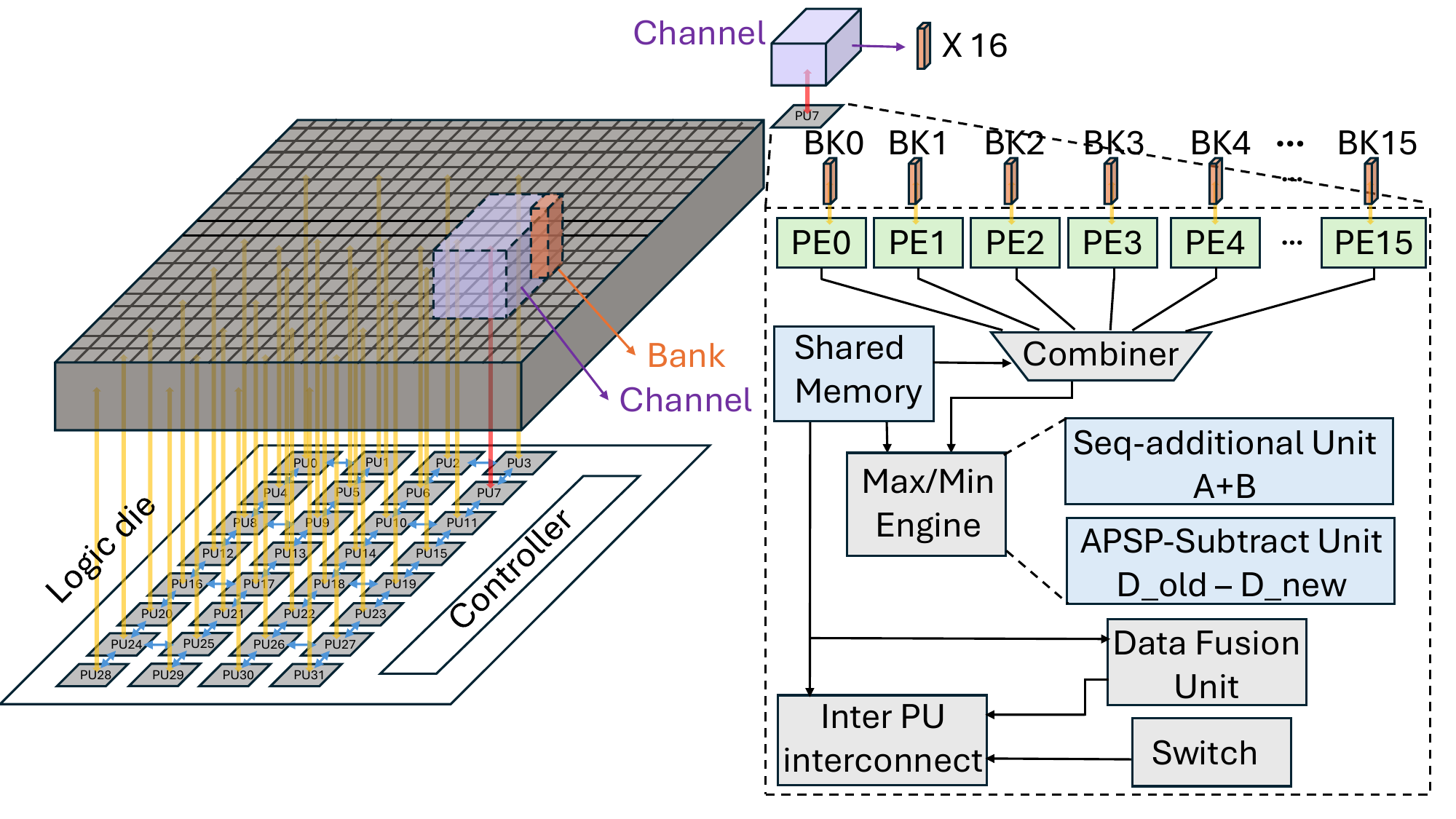}
    \caption{Overview of the GenDRAM System Architecture. The left panel shows the Logic Die connected to the 3D DRAM stack, with an array of PU. The right panel details the internal structure of a Near Bank PE Cluster, highlighting its PEs and specialized computational units for APSP and Sequence Alignment}
    \label{fig:gendram_overview_arch}
\end{figure}

GenDRAM is a novel 3D PIM accelerator designed to address the memory-bound challenges of both genomic analysis and APSP graph processing workloads. Leveraging the advanced capabilities of M3D DRAM, GenDRAM tightly integrates a high-performance Near-Memory Processor (NMP) on a dedicated logic die with a multi-layered DRAM stack. This architecture, broadly depicted in Fig. \ref{fig:gendram_overview_arch}, aims to eliminate the traditional memory wall bottleneck by bringing computation directly to the data, facilitating massive internal bandwidth and reducing data movement.

As shown on the left side of Fig. \ref{fig:gendram_overview_arch}, the GenDRAM system comprises a multi-layered M3D DRAM stack directly bonded to an advanced-node logic die. The logic die houses an array of numerous PUs, organized into clusters, along with a central controller. These PUs are strategically placed near the DRAM banks, connected via dedicated bank channels. This proximity and high-bandwidth connection are crucial for enabling efficient PIM operations. Data access is initiated when a PU's local memory controller issues an \texttt{ACTIVATE} command to a bank. This action latches an entire DRAM row into the bank's Row Buffer. The PU's internal computational units then read their necessary data directly from this wide row buffer over the high-bandwidth hybrid bonding interface, performing in-situ computation before the row is precharged. The overall architecture emphasizes a modular and scalable design, where each PU is capable of operating independently or collaboratively with others, coordinated by the central Controller. The entire system is then connected to a host CPU.

\subsection{M3D DRAM-Logic Co-Optimization}
\label{sec:co_optimization}

The microarchitecture of the Logic Die is co-designed to saturate the massive internal bandwidth of the M3D DRAM. We specifically optimize the PU count to match the 32 independent DRAM bank-groups and configure the I/O width to leverage the dense hybrid bonding interface, balancing maximum throughput against silicon area limitations.

\subsubsection{Bandwidth-Matched PU Scaling}
The M3D DRAM is organized into 32 independent Bank-Groups (derived from 16 Channels $\times$ 2 Groups/Channel) to maximize vertical parallelism. To fully saturate the massive internal bandwidth ($\sim$34 TB/s) enabled by hybrid bonding, we establish a 1:1 coupling ratio between DRAM Bank-Groups and compute/search PUs. Consequently, the logic die implements exactly 32 PUs. Increasing PUs beyond this count would result in memory contention, while fewer PUs would leave the internal bandwidth underutilized. More details can be found in Section ~\ref{PUPE_tradeoff}

\subsubsection{Hybrid Bonding Interface}
Unlike conventional HBM, which is limited by 1024-bit TSV interfaces per stack, the Cu-Cu hybrid bonding enables a dense vertical interconnect pitch ($\sim 1 \mu m$). We configure each PU with a dedicated 1024-bit wide I/O interface directly connected to its vertically aligned DRAM Bank-Group. This allows a single \texttt{ACTIVATE} command to fetch an entire vector (e.g., 32 integers $\times$ 32-bit) into the PU's shared memory in a single cycle, perfectly matching the SIMD width required by our compute PEs for tile processing.

\subsection{Process Units}

The right panel of Fig.~\ref{fig:gendram_overview_arch} provides a detailed view of the PU, also referred to as the near bank PE cluster, which serves as the fundamental computational block within the GenDRAM architecture. Each PU is strategically positioned on the logic die to be directly associated with a specific set of underlying DRAM banks (illustrated as BANK 0 through BANK 15), granting it high-bandwidth, low-latency access to its designated local data partition within the M3D DRAM stack. Internally, the PU comprises several key components orchestrated around a central shared memory. This includes a cluster of 16 lightweight PEs (PE 0 through PE 15), which act as the primary engines for parallel computation. Data flows from the associated DRAM banks into the shared memory, which then serves as a high-speed scratchpad and data distribution hub for the PEs and other specialized execution units within the PU.

Integral to the PU's versatility is a dedicated Max/Min Engine, specifically designed to accelerate the core arithmetic patterns common to both APSP and sequence alignment. This engine is not a monolithic block but rather a collection of specialized sub-units accessible via the shared memory. It includes an APSP-subtract unit tailored for the difference calculations ($D_{old} - D_{new}$) often used in graph algorithms, a Seq-additional Unit optimized for the simple additions ($A+B$) prevalent in sequence alignment scoring, and a generic max/min Unit for performing the fundamental $\max(A)$ or $\min(A)$ selections required by DP relaxation steps. This flexible hardware allows GenDRAM to efficiently execute diverse DP tasks without the overhead of general-purpose multipliers. Furthermore, each PU incorporates an inter-PU interconnect interface, featuring a data fusion unit and switch. This enables efficient communication and data aggregation between different PUs across the logic die, facilitating the complex data exchanges required by blocked algorithms like FW and the pipelined stages of the genomics workflow.

\subsection{Process Elements}
To efficiently handle the diverse computational requirements of genomic analysis and graph processing, GenDRAM employs a heterogeneous PU design. In this research, the 32 PUs on the logic die are specialized into two types: search and compute PUs. Specifically, 8 PUs are dedicated as search PUs, while the remaining 24 PUs function as compute PUs. Each type of PU contains multiple specialized PEs tailored for their respective tasks.

\begin{figure}[htbp]
    \centering
    \includegraphics[width=0.8\columnwidth]{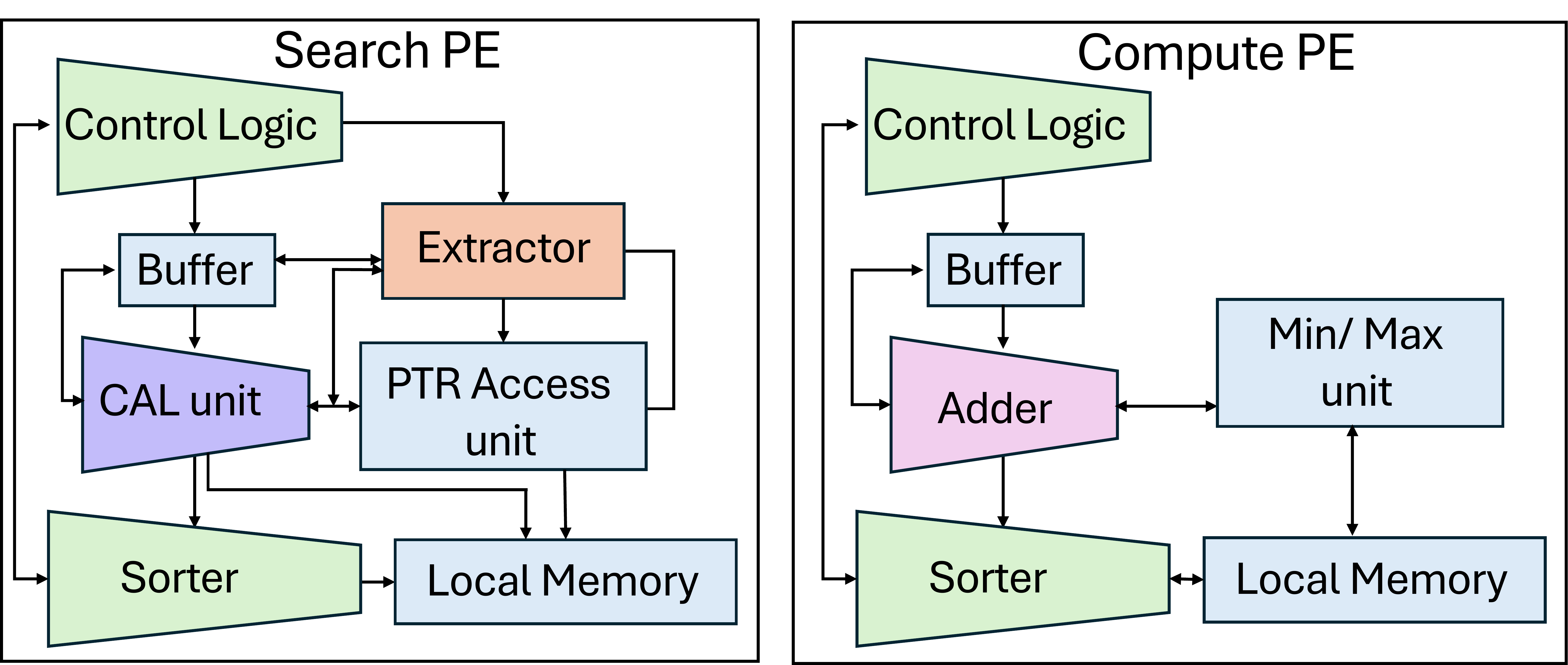}
    \caption{Internal Architecture of Search PE (left) and Compute PE (right). Search PEs are optimized for rapid pattern matching and data extraction for the Seeding phase, while Compute PEs are designed for arithmetic and comparison operations crucial for dynamic programming and graph algorithms}
    \label{fig:pe_internal_structure}
\end{figure}

\textbf{Search PE - for seeding:} Illustrated on the left side of Fig. \ref{fig:pe_internal_structure}, search PEs are specifically designed to accelerate the memory-intensive \textit{seeding} phase of genomic alignment.  The \textit{seeding phase} is characterized by irregular, latency-critical memory accesses and a high demand for data filtering. This memory-bound property makes it an ideal candidate for near-bank acceleration to filter massive datasets at the source. Consequently, the Search PE is designed with specialized logic to handle these properties: a pointer table(PTR) qccess unit and candidate alignment location(CAL) unit to accelerate the non-sequential table lookups ~\cite{khaleghi2022salient}, and an extractor and sorter to perform filtering. This design transforms the massive raw bandwidth of M3D DRAM into a stream of high-quality candidate seeds, mitigating the bottleneck of transferring redundant data.

\textbf{Compute PE - for alignment \& APSP:} Shown on the right side of Fig. \ref{fig:pe_internal_structure}, the alignment and APSP phases are compute-bound with high data locality. Both workloads rely on iterative updates (wavefront or pivot-based) where operands are reused extensively among neighboring cells. 
    To exploit this regularity, the Compute PE functions as a lightweight arithmetic engine. It features a local memory to capture temporal data locality and minimize DRAM activation overhead, coupled with a reconfigurable adder/comparator to execute the core semi-ring operations ($\min/+$ or $\max/+$) efficiently. This architecture leverages the massive parallelism of the PU array to handle the cubic complexity of DP workloads.

\section{M3D DRAM Mapping and Execution}
\label{sec:mapping_execution}

The performance of GenDRAM relies on a co-designed strategy that exploits the physical geometry of M3D DRAM. We decompose this into two orthogonal optimizations: \textit{Spatial optimization}, which maps data structures to specific DRAM tiers and channels based on latency and bandwidth requirements, and \textit{temporal optimization}, which dynamically reconfigures the dataflow between Processing Units (PUs) to match the algorithmic structure of the workload.

\subsection{Spatial optimization: 3D-Aware Data Mapping}
\label{ssec:spatial_mapping}

GenDRAM exploits the unique physical characteristics of M3D DRAM—specifically, the vertical latency variation and the massive horizontal parallelism—to optimize conflicting memory requirements.

\subsubsection{Latency-critical mapping (tiering)}
\label{ssec:alignment_mapping}
As established in prior work on M3D DRAM ~\cite{pan2025stratum}, vertical layers exhibit linear latency variation due to the staircase wordline structure. Layers closer to the logic die have significantly lower $t_{RCD}$ compared to top layers. 

For the seeding phase of the genomics pipeline, memory access latency is the primary bottleneck, often accounting for 60\%--80\% of the total execution time in standard software aligners due to excessive pipeline stalls~\cite{xu2023rapidx}. The seeding mechanism involves a dependent two-stage lookup: first accessing the Pointer (PTR) table to find the bucket index, and subsequently retrieving coordinates from the Candidate Location (CAL) table ~\cite{khaleghi2022salient}. This dependency chain creates a simplified random-access pattern that stalls execution if memory latency is high. To mitigate this, we map these latency-sensitive PTR and CAL tables (totaling $\sim$17GB) to the bottom-most Tier (Tier 0), which comprises the physical layers closest to the logic die. In our 1024-layer stack organized into 8 latency tiers, Tier 0 offers the lowest $t_{RCD}$ (2.29 ns).

\subsubsection{Bandwidth-critical mapping (interleaving)}
For compute-bound tasks like APSP and genomic alignment, aggregate memory bandwidth is paramount. In this research, we employ a channel-interleaved mapping strategy to maximize parallelism across all the Compute PUs.

\begin{figure}[htbp]
    \centering
    \includegraphics[width=\columnwidth]{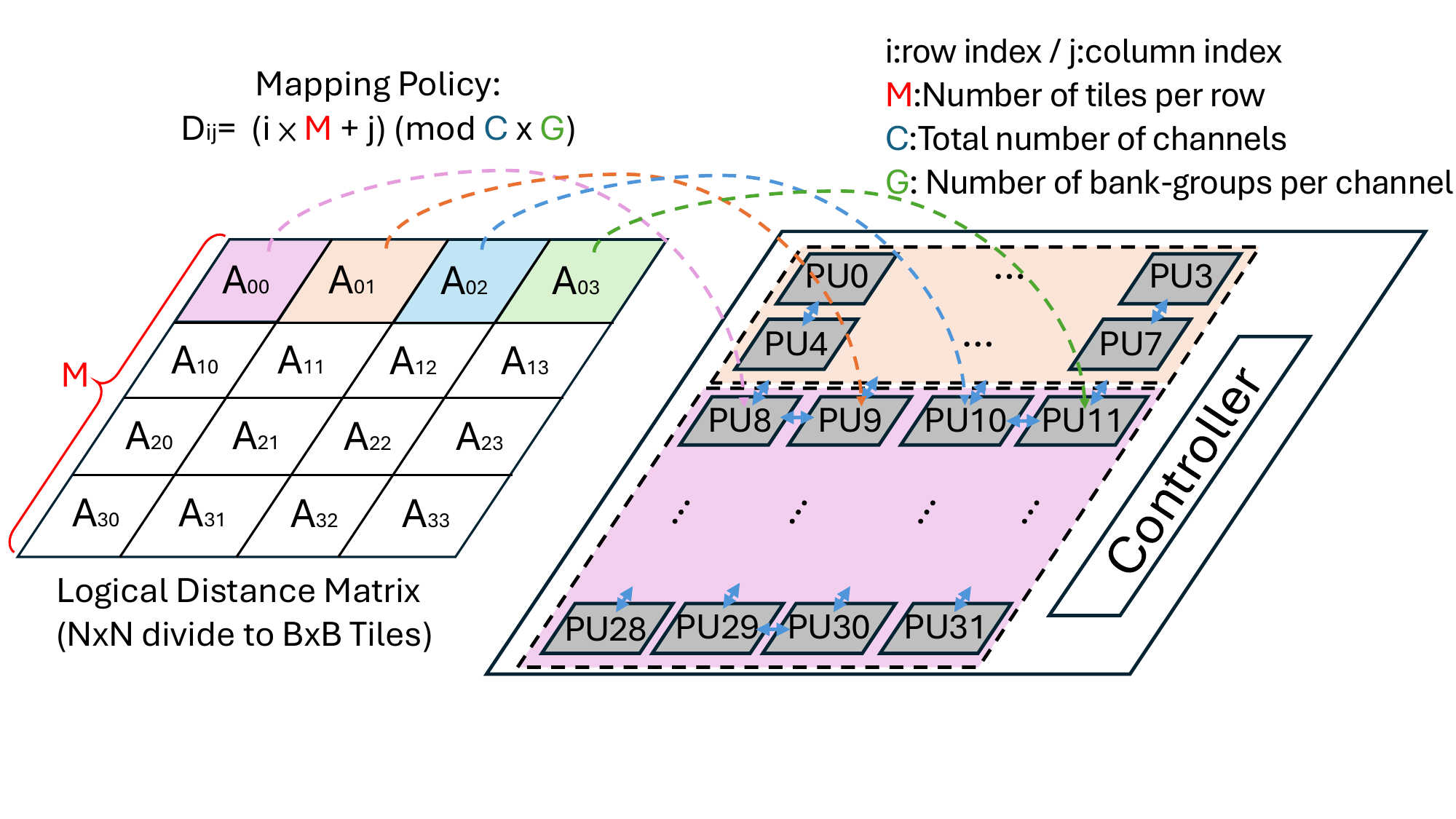}
    \caption{Illustration of the Logical-to-Physical Mapping Mechanism for APSP. The distance matrix is logically partitioned into tiles and distributed across the Compute PUs using a modulo-based policy shown in the equation. This mechanism ensures that logically adjacent tiles are mapped to distinct PU, enabling conflict-free parallel access}
    \label{fig:apsp_interleaved_mapping}
\end{figure}

As illustrated in Fig.~\ref{fig:apsp_interleaved_mapping}, the mapping mechanism translates the logical coordinates of a matrix tile $(i, j)$ directly to a physical Compute PU ID. The mapping policy is defined by:
\begin{equation}
    \text{Target PU} = (i \times M + j) \pmod{C \times G}
\end{equation}
where $M$ is the number of tiles per row, $C$ is total number for the channel and $G$ is number of bank-group per channel. This modulo-based distribution ensures that logically adjacent tiles—such as $A_{00}$ (pink) and $A_{01}$ (orange) in the figure—are mapped to distinct physical PUs (PU0 and PU4, respectively). Consequently, during wavefront propagation where neighbor tiles are accessed simultaneously, requests are naturally distributed across groups of banks and PUs. This eliminates bank conflicts and allows the system to fully saturate the M3D DRAM's internal bandwidth.

\subsection{Temporal optimization: reconfigurable execution flow}
\label{ssec:temporal_scheduling}

The logic die controller dynamically reconfigures the interactions between PUs to match the distinct dataflow patterns of the target applications.

\subsubsection{Mode 1: Homogeneous systolic broadcast (APSP)}
\begin{figure}[htbp]
    \centering
    \includegraphics[width=\columnwidth]{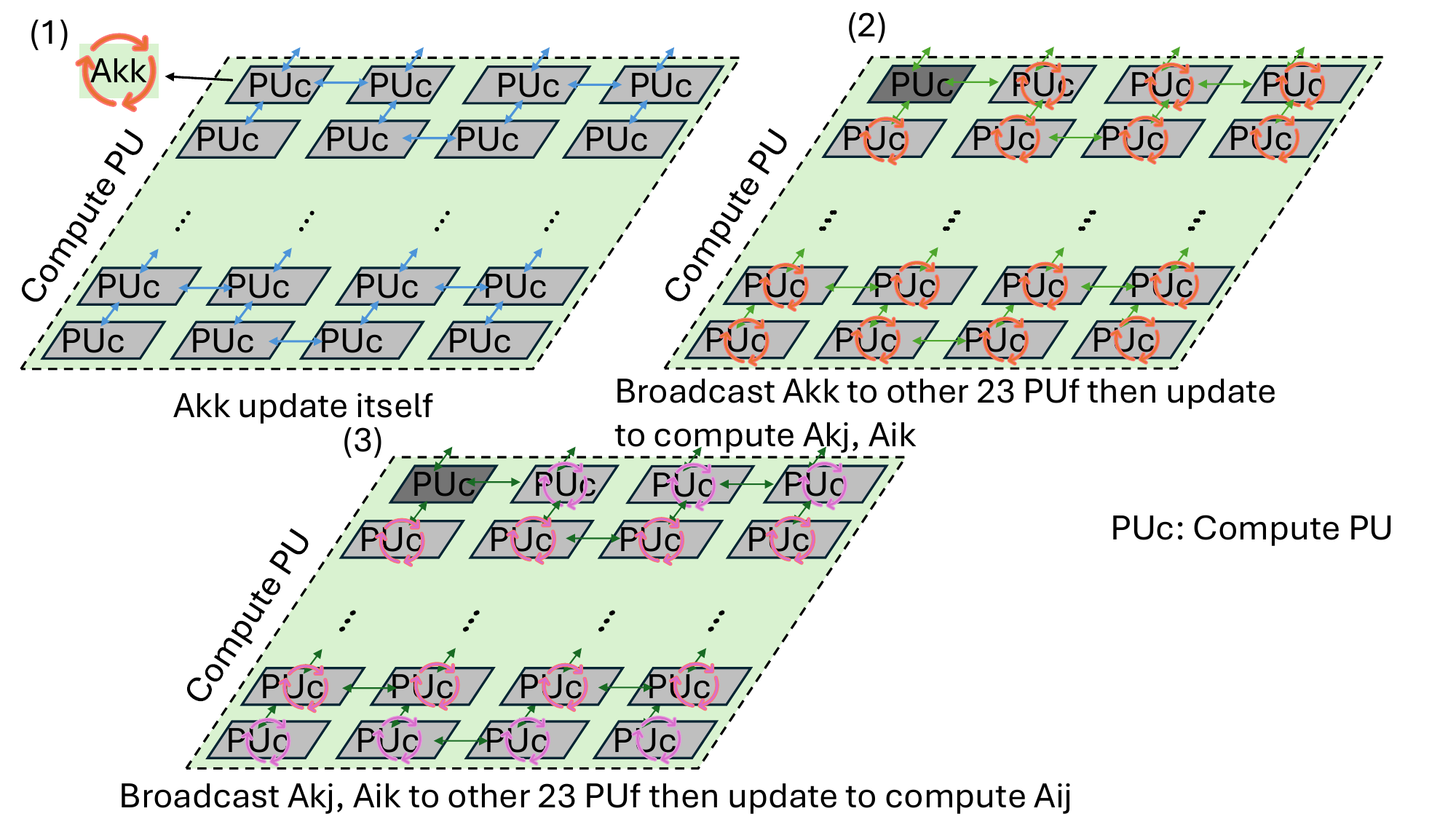}
    \caption{Homogeneous scheduling for APSP. The interconnect facilitates broadcasting of pivot data to synchronize parallel updates across $N_{comp}$ Compute PUs}
    \label{fig:APSPscheduling}
\end{figure}

For APSP, the workload is uniform and compute-bound. GenDRAM configures the PU array into a \textit{homogeneous parallel mode}, utilizing the set of $N_{comp}$ specialized Compute PUs. The inter-PU interconnect functions as a broadcast network to support the Blocked FW algorithm. The execution of each super-step proceeds in three synchronized phases (Fig.~\ref{fig:APSPscheduling}):
\begin{enumerate}
    \item \textbf{Pivot update:} A single PU updates the pivot block $A_{kk}$.
    \item \textbf{Row/Column broadcast:} The updated pivot block is broadcast via the Ring Router (128 GB/s) to all other PUs. This allows $N_{comp}$ PUs to update their respective pivot row/column blocks in parallel.
    \item \textbf{Systolic update:} Updated row/column blocks are propagated to all PUs. The array operates as a distributed systolic system, performing the $O(N^3)$ internal updates ($A_{ij}$) in parallel without stalling for memory access.
\end{enumerate}

\subsubsection{Mode 2: Heterogeneous Pipelining (Genomics)}
\begin{figure}[htbp]
    \centering
    \includegraphics[width=\columnwidth]{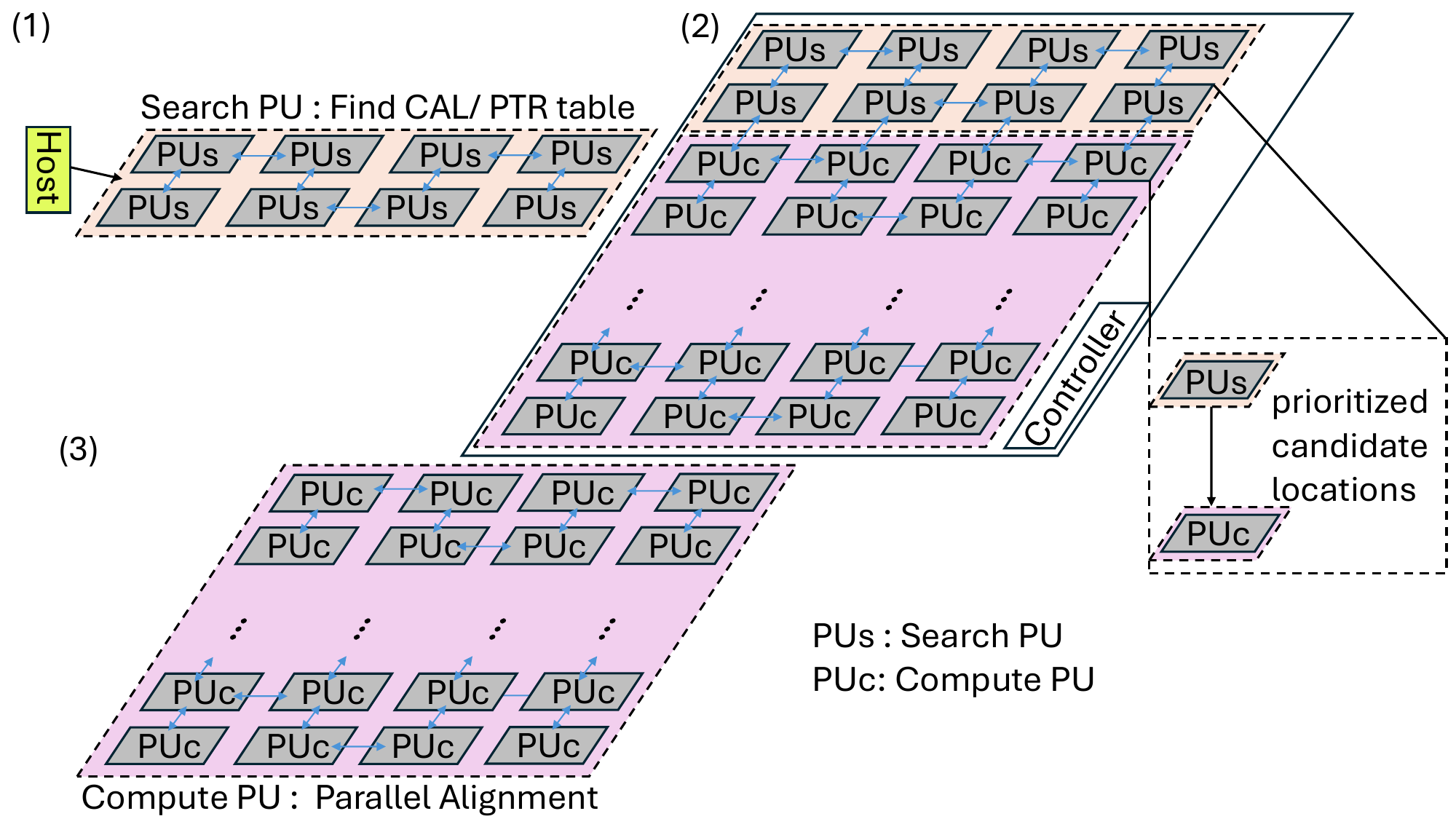}
    \caption{Heterogeneous pipeline Scheduling for genomics. Search PUs (producer) stream candidate locations to Compute PUs (consumer), overlapping memory-bound seeding with compute-bound alignment}
    \label{fig:bioscheduling}
\end{figure}

For genomics, the workload is composed of distinct memory-bound (Seeding) and compute-bound (Alignment) stages. GenDRAM reconfigures the array into a \textit{Heterogeneous pipeline mode} (Fig.~\ref{fig:bioscheduling}), partitioning the resources into $N_{search}$ Search PUs and $N_{comp}$ Compute PUs:
\begin{itemize}
    \item \textbf{Producer stage (Search PUs):} The $N_{search}$ Search PUs operate as a high-throughput front-end. They continuously fetch reads and perform latency-sensitive index lookups in the fast DRAM tiers to generate candidate locations.
    \item \textbf{Consumer stage (Compute PUs):} The $N_{comp}$ Compute PUs act as the back-end. They receive candidate metadata via the interconnect and perform the compute-heavy banded alignment.
\end{itemize}
This pipelined handoff effectively decouples the two stages. By balancing the throughput of $N_{search}$ Search PUs with $N_{comp}$ Compute PUs, the latency of the memory-bound seeding phase is entirely hidden behind the alignment computation, ensuring high utilization of the compute resources.

\section{Evaluation}
\label{sec:evaluation}
Our comprehensive evaluation,detailed in this section, demonstrates that GenDRAM delivers large benefits in performance and efficiency while adhering to strict physical constraints. Specifically, it achieves up to \textbf{67$\times$} and \textbf{22$\times$} speedup for APSP and genomic alignment, respectively, compared to the NVIDIA A100 GPU. This high throughput is achieved within a sustainable thermal envelope. Simulation results indicate an average power consumption of only \textbf{31.2 W} for the complex bioinformatics pipeline and \textbf{10.2 W} for APSP. These values are well within the thermal budget of passively cooled 3D stacks, validating the physical feasibility of our monolithic design. GenDRAM also has superior energy efficiency, outperforming the A100 baseline by \textbf{152$\times$} and the domain-specific RapidGraph~\cite{chen2025rapid} accelerator by \textbf{20$\times$}. The following subsections detail our methodology, followed by an in-depth analysis of performance, energy efficiency and area constraints, sensitivity \& scalability analysis, followed by the discussion on overhead.

\subsection{Experimental setup}

\subsubsection{\textbf{Datasets}}
We evaluate the architecture using two distinct sets of workloads. For the APSP evaluation, three real-world graph datasets were selected from the SNAP collection~\cite{leskovecjunSNAP} and OpenStreetMap~\cite{mooney2017reviewOSM} to ensure robustness across diverse topologies and scales. These include the \texttt{ca-GrQc} collaboration network ($N$=5,242) and the \texttt{p2p-Gnutella08} peer-to-peer network (p2p)($N$=6,301)~\cite{leskovecjunSNAP}, alongside a significantly larger road network graph, Open Street Map (OSM)($N$=65,536)~\cite{mooney2017reviewOSM}. This selection facilitates a comprehensive scalability analysis of GenDRAM against conventional accelerators on graphs with fundamentally different structures.

For genomic sequence alignment, all synthetic reads are aligned against the GRCh38 human reference genome~\cite{GRCh38}. Short-read datasets are generated using the Mason simulator~\cite{holtgrewe2010mason} to mimic Illumina reads with a 5\% error rate. Long-read datasets are generated using the PBSIM simulator~\cite{ono2013pbsim} to model high-error-rate technologies, specifically PacBio reads with a 15\% error rate and ONT reads with a 30\% error rate.

\subsubsection{\textbf{Measured Baselines}}
We compare our simulated results to physical measurements of both APSP workloads, while we ran Minimap2~\cite{li2018minimap2, intelpennycook2013exploring} and GASAL2~\cite{ahmed2019gasal2} aligners by measuring on an Intel Xeon CPU and an NVIDIA A100 GPU~\cite{choquette2021nvidiaA100}. Furthermore, we include projected performance for the NVIDIA H100, where estimates are analytically extrapolated based on the architectural bandwidth and compute throughput scaling factors reported in detailed benchmarking studies~\cite{luo2402benchmarkinggraphH100, samarakoon2025benchmarkingH100}, representing state-of-the-art GPU with 2.0 TB/s HBM2e bandwidth running GASAL2 graph algorithms~\cite{ahmed2019gasal2} and minimap2 sequence aligner~\cite{li2018minimap2, guo2019hardwareFPGA}. 


\subsubsection{\textbf{Simulated Accelerator Baselines}}
\textit{Here you list all the accelerators you compared to for APSP, short and long reads.}

\subsubsection{\textbf{GenDRAM Configuration \& Evaluation}}
\label{ssec:exp_setup}
GenDRAM features a 32 GB, 1024-layer M3D DRAM stack integrated with a 7nm logic die~\cite{pan2025stratum}. As detailed in Table~\ref{tab:3ddram_params}, the memory system utilizes a 35 nm feature size and is organized into 16 channels per chip with a high-bandwidth interface mirroring HBM3. Crucially, the stack is partitioned into 8 tiers, each providing 4 GB of storage with varying row-to-column delays ($t_{\text{RCD}}$). This tiered structure is fundamental to the latency optimization strategy discussed in Section~\ref{ssec:alignment_mapping}, allowing latency-critical seeding data to be mapped to faster physical layers.   

The GenDRAM logic processor, comprising 32 PUs (8 Search and 24 Compute) operating at 1~GHz, is modeled in Verilog and synthesized using Synopsys design compiler on a 45nm process node~\cite{xu2023rapidx}, with metrics subsequently scaled to a 7nm technology node similar to~\cite{pan2025stratum}. The computing fabric utilizes a 32-bit fixed-point format for APSP while retaining compatibility with 5-bit operations for difference-based alignment~\cite{xu2023rapidx}, supported by on-chip SRAMs characterized using FinCACTI~\cite{shafaei2014fincacti}.

\begin{table}[h]
\centering
\caption{GenDRAM M3D DRAM Device and System Parameters (Final Stable Structure)}
\label{tab:3ddram_params}
\begin{tabular}{|c|c|c|c|} 
\hline
\multicolumn{4}{|c|}{\textbf{M3D DRAM Device Parameter}} \\
\hline
Layers & 1024 & Feature Size & 35 nm \\
\hline
BL/WL Pitch & 70 nm/1 um & Staircase Pitch & 500 nm \\
\hline
MAT Size & $1\text{k} \times 1\text{k}$ & \#MATs/Bank & $32 \times 32$ \\
\hline
Bank Storage & 1 Gb & Bank Area & 0.44 mm$^2$ \\
\hline
Row Buffer & 32 Kb & Energy/bit & 0.429 pJ \\
\hline
Chip Area & 121 mm$^2$ & Chip Storage & 32 Gb \\
\hline
\multicolumn{4}{|c|}{\textbf{M3D DRAM System Parameters}} \\
\hline
Tier Design & \multicolumn{3}{|c|}{8 tiers; 4Gb storage per tier} \\
\hline
Organization & \multicolumn{3}{|c|}{\makecell[c]{16 Channels per chip (64b data I/O per channel); \\ 16 banks per channel.}} \\
\hline
DRAM Timing & \multicolumn{3}{|c|}{
    \makecell[l]{
        $t_{\text{RCD}}$ = [2.29, 3.92, 5.99, 8.50, 11.44, 14.82, 18.63, \\ 
        \quad  22.88] ns; \\
        $t_{\text{RP}}$ = 4.77 ns;  $t_{\text{RAS}}$ =  $t_{\text{RCD}}$ + 27.5 ns; \\
        $t_{\text{RC}}$ = $t_{\text{RP}}$ + $t_{\text{RAS}}$
    }
} \\
\hline
CPU-DRAM I/F & \multicolumn{3}{|c|}{1024b data I/Os; 6.4 Gbps per pin (same as HBM3)} \\
\hline
\end{tabular}
\end{table}

\textbf{GenDRAM evaluation} is done using a cycle-accurate simulator. The functional behavior of the core PEs is implemented in Verilog. To derive realistic physical parameters, the on-chip SRAM components, such as the shared memory within each PU, are modeled using FinCACTI ~\cite{shafaei2014fincacti}, calibrated with publicly available 7nm SRAM data. The area and energy metrics for the specialized, multiplier-less Compute PEs and the pipeline-based Search PEs are estimated based on synthesis reports scaled to 7nm ~\cite{pan2025stratum, xu2023rapidx}.

To evaluate the performance, power, and area (PPA) of our proposed GenDRAM accelerator, we developed a detailed, cycle-accurate simulator derived based on previous work~\cite{pan2025stratum}. The simulator models our heterogeneous NMP architecture, including the multi-level scheduling and data mapping strategies discussed previously. PPA metrics are rigorously evaluated through a multi-stage methodology: area is extracted from synthesis reports, energy is quantified via post-synthesis simulations with annotated switching activity, and end-to-end execution cycles are determined by an in-house cycle-accurate Python simulator that models detailed data mapping and scheduling policies.

Energy consumption is determined through the simulations, which track the activity of each hardware component. The energy metrics for fundamental operations (e.g., a 16-bit addition) and memory accesses are derived from literature and synthesis reports, and the total energy is accumulated based on the switching activity for a given workload. Execution cycles, on-chip communication cycles, and associated latency metrics are derived directly from the in-house simulator. The simulator is designed to be highly configurable, taking as input the graph size and structure for APSP, or the batch of query reads and their lengths for genomic alignment. It also models the detailed data mapping and the heterogeneous pipeline scheduling policies discussed in Section \ref{sec:mapping_execution}. As output, the simulator provides the overall end-to-end execution time, as well as detailed PPA breakdowns at the component level for each application, enabling a comprehensive analysis of the GenDRAM architecture.

\subsection{GenDRAM APSP performance and energy efficiency}
\label{sec:perf}

\subsubsection{GenDRAM APSP performance}

\begin{figure}
  \centering
  \includegraphics[width=1.0\columnwidth]{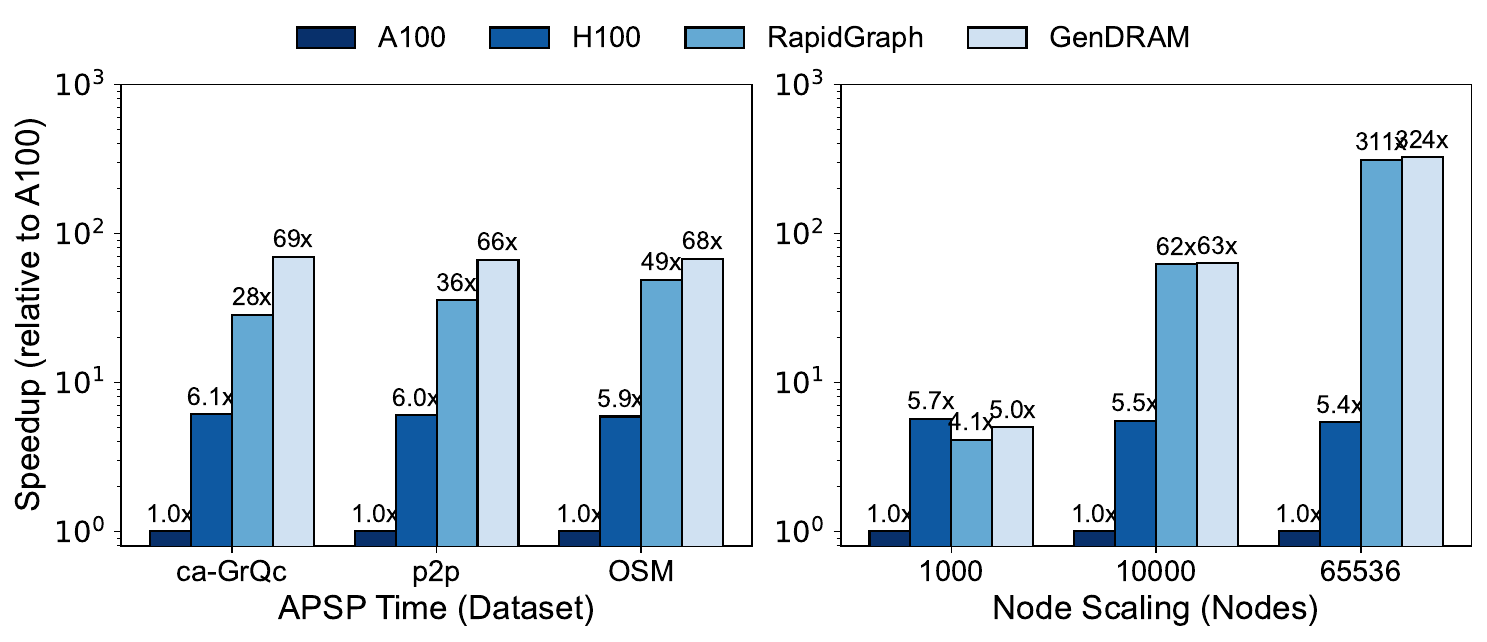}   
  \caption{GenDRAM performance comparison for APSP}
  \label{fig:apsp_perf}
\end{figure}

We conducted a series of simulations using the blocked FW algorithm to evaluate the performance of the GenDRAM accelerator for APSP problems.  Fig.~\ref{fig:apsp_perf} shows comparison of GenDRAM, with measurements on NVIDIA A100 GPU and estimates on H100 and the ReRAM-based PIM accelerator RapidGraph~\cite{chen2025rapid}. 

The results demonstrate a transformative performance improvement for GenDRAM. As shown in the left panel of Fig.~\ref{fig:apsp_perf}, on the large-scale OpenStreetMap (OSM) dataset, GenDRAM achieves a staggering \textasciitilde68$\times$ speedup over the A100 and a \textasciitilde11.3$\times$ speedup over the H100. This massive gain is a direct consequence of the PIM architecture resolving the memory wall bottleneck, preventing the stall cycles that keep GPU cores idle for over 60\% of execution time~\cite{sao2021scalableGPU}. Moreover, GenDRAM outperforms RapidGraph ,which achieves $\sim$49$\times$, by approximately 1.4$\times$. This advantage stems from the underlying memory technology: while ReRAM-based solutions like RapidGraph suffer from high write latency during the frequent $D_{ij}$ updates inherent to APSP, GenDRAM's DRAM-based substrate handles these write-intensive operations with significantly lower latency.

The right panel of Figure~\ref{fig:apsp_perf} illustrates how this performance gap scales with problem size. As the number of nodes increases from 1,000 to 65,536, the advantage of GenDRAM widens dramatically, reaching a peak speedup of \textasciitilde324$\times$ over the A100 at N=65,536. Notably, while RapidGraph also scales well ($\sim$311$\times$), GenDRAM maintains a consistent lead. This validates that GenDRAM is not only exceptionally fast but also highly scalable, leveraging the massive internal bandwidth of M3D DRAM to efficiently handle the $O(N^3)$ computational load without the write-penalty bottlenecks of non-volatile memories.

\subsubsection{GenDRAM APSP energy efficiency}

\begin{figure}[htbp]
    \centering
    \includegraphics[width=1.0\columnwidth]{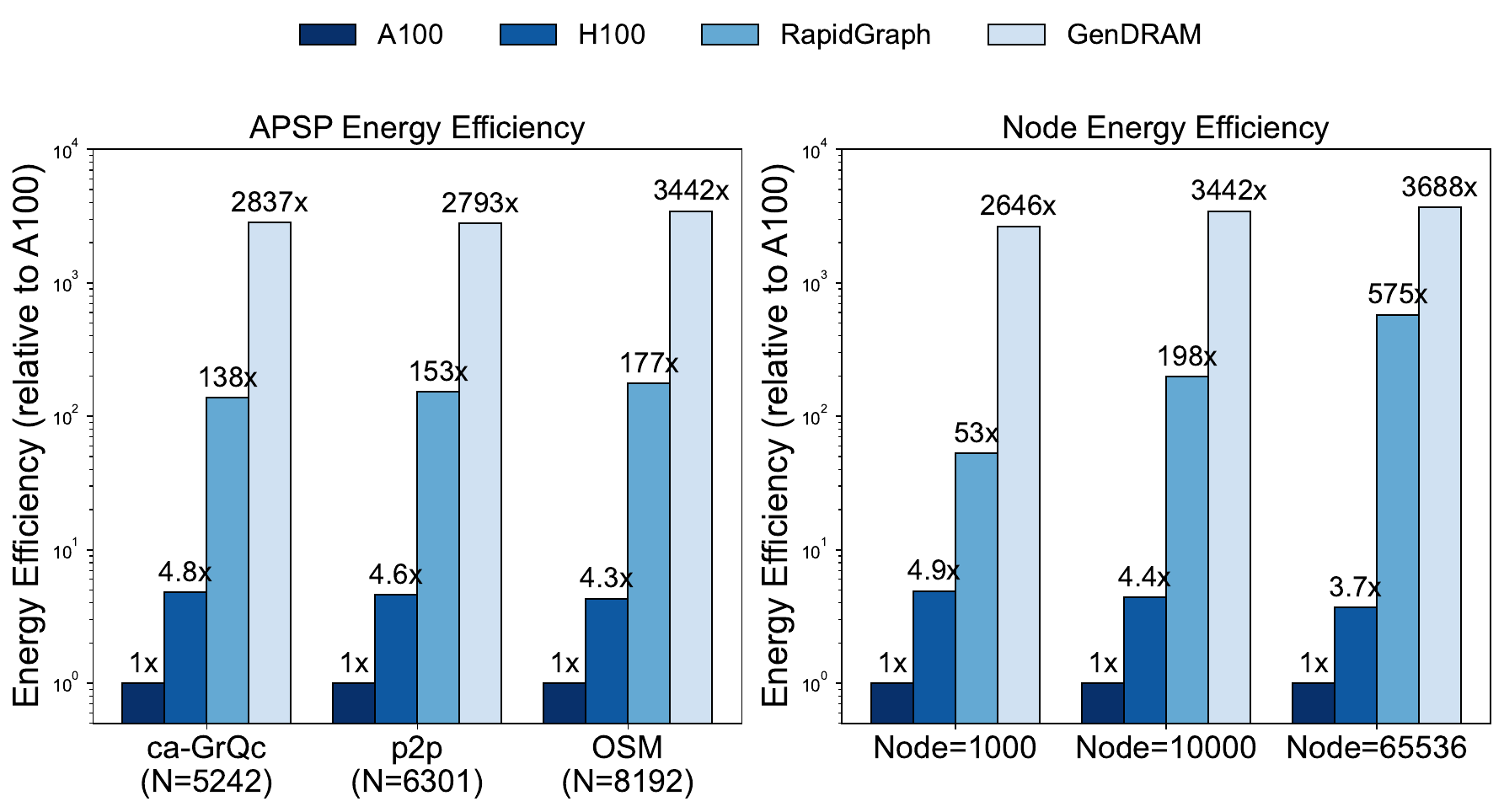}
    \caption{Energy efficiency for APSP under different node dataset}
    \label{fig:APSP_EE}
\end{figure}
We evaluated GenDRAM's energy efficiency on the $O(N^3)$ APSP algorithm, with results shown in Fig. \ref{fig:APSP_EE}. Normalized to the A100 GPU baseline, GenDRAM achieves transformative efficiency improvements ranging from 2,837$\times$ for \texttt{ca-GrQc} to 3,442$\times$ for \texttt{OpenStreetMap}, maintaining this advantage as the problem size scales to 3,688$\times$ at N=65536. This performance stands in stark contrast to the estimated H100 baseline, which only provides 3.7$\times$ to 4.9$\times$ improvement. Furthermore, compared to the ReRAM-based PIM accelerator RapidGraph ~\cite{chen2025rapid}, which achieves notable gains of 138$\times$ to 575$\times$ over the A100, GenDRAM still delivers an approximately 20$\times$ higher efficiency. This divergence stems from fundamental architectural characteristics: while GPUs are bottlenecked by the high energy cost of off-chip data movement, RapidGraph is constrained by the significant energy penalty of \textit{writing} to ReRAM devices and the overhead of analog-to-digital conversion required for write-intensive APSP updates. GenDRAM achieves the highest efficiency by combining low-power digital logic PEs with 3D DRAM, effectively eliminating off-chip traffic entirely while avoiding the write-energy penalties inherent to ReRAM technologies.

\subsection{GenDRAM alignment performance and energy efficiency}
\textbf{GenDRAM sequence alignment performance:}
To provide a comprehensive evaluation, we benchmark GenDRAM against a diverse set of baselines. All performance results are normalized to Minimap2 ~\cite{li2018minimap2}, a widely-used software aligner running on an Intel Xeon CPU MAX 9460 (8 cores), which serves as the 1.0$\times$ baseline. For hardware acceleration comparisons, we utilize the NVIDIA A100 and H100 GPUs running state-of-the-art libraries: GASAL2~\cite{ahmed2019gasal2} for short reads and minimap2~\cite{guo2019hardwareFPGA} for long reads. Additionally, we include domain-specific accelerators, including the ASIC-based ABSW~\cite{liao2018adaptivelyABSW} and the ReRAM-based RAPIDx~\cite{xu2023rapidx}, to show GenDRAM's performance against specialized hardware.  

The end-to-end performance of GenDRAM was evaluated using the GRCh38 human reference genome, separating execution into memory-intensive seeding and compute-intensive alignment phases. While the NVIDIA A100 baseline is  constrained by RAM latency during seeding, accounting for two-thirds of execution time, GenDRAM successfully removes this bottleneck. By leveraging dedicated Search PUs and 3D-aware tiered memory, GenDRAM accelerates seeding by approximately 138$\times$, shifting the critical path to alignment, where specialized Compute PEs still deliver an 8.5$\times$ speedup. Consequently, GenDRAM achieves an aggregate performance improvement of over 22$\times$ relative to the A100 and 11$\times$ vs the H100.

\begin{figure}
  \centering
  \includegraphics[width=1.0\columnwidth]{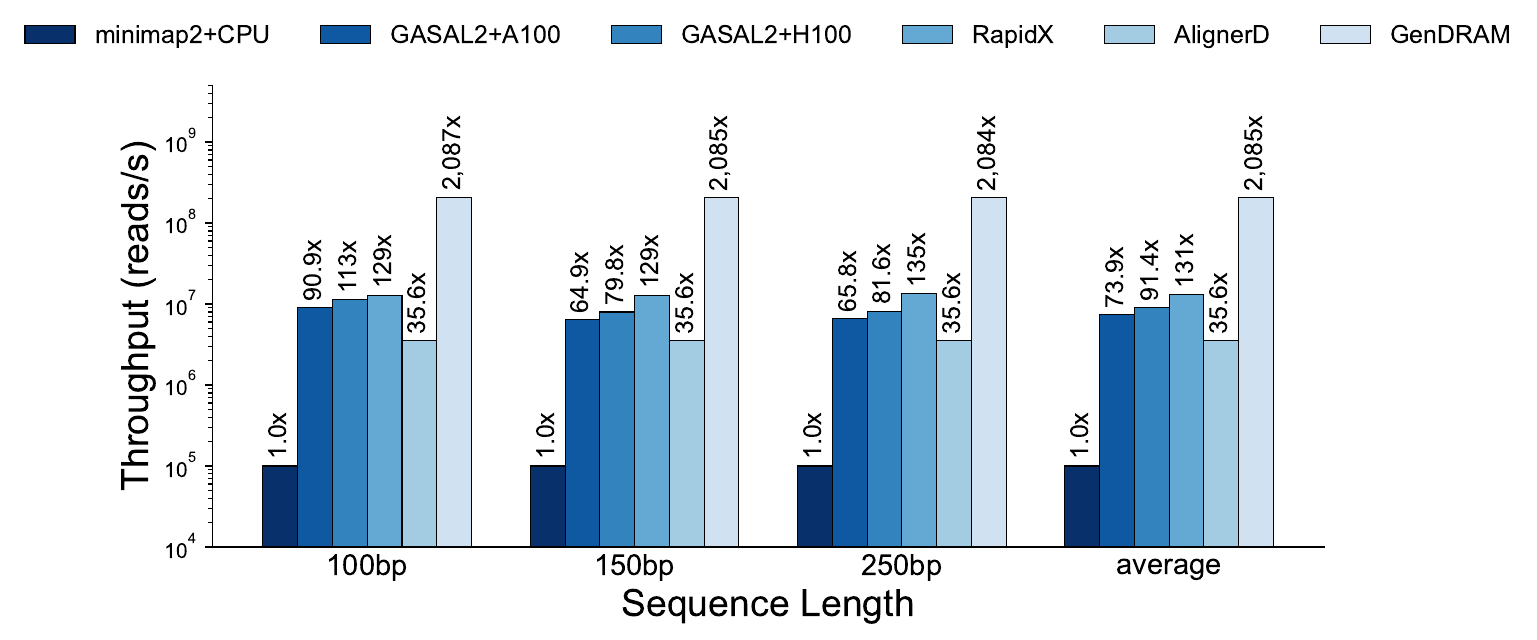}    
  \caption{Alignment throughput comparison of GenDRAM, RAPIDx~\cite{xu2023rapidx}, GASAL2~\cite{ahmed2019gasal2}, AlignerD~\cite{zhang2023alignerd} and Minimap2 acceleration~\cite{guo2019hardwareFPGA} for short reads}
  \label{fig:genomic_perf1}
\end{figure}

Performance on short reads, illustrated in Fig.~\ref{fig:genomic_perf1}, shows that relative to the CPU baseline, GenDRAM achieves orders-of-magnitude improvement. When compared to the highly optimized GPU baselines, it achieves average throughputs approximately 45$\times$ and 23$\times$ higher than the A100 (running GASAL2) and H100, respectively. This significant lead results from the inefficiency of general-purpose GPU architectures in handling latency-sensitive, high-overhead short-read workloads. GenDRAM resolves these limitations through its fully integrated pipeline, where dedicated Search PUs eliminate the seeding memory bottleneck and streamlined Compute PEs efficiently handle alignment. 

\begin{figure}
  \centering
  \includegraphics[width=1.0\columnwidth]{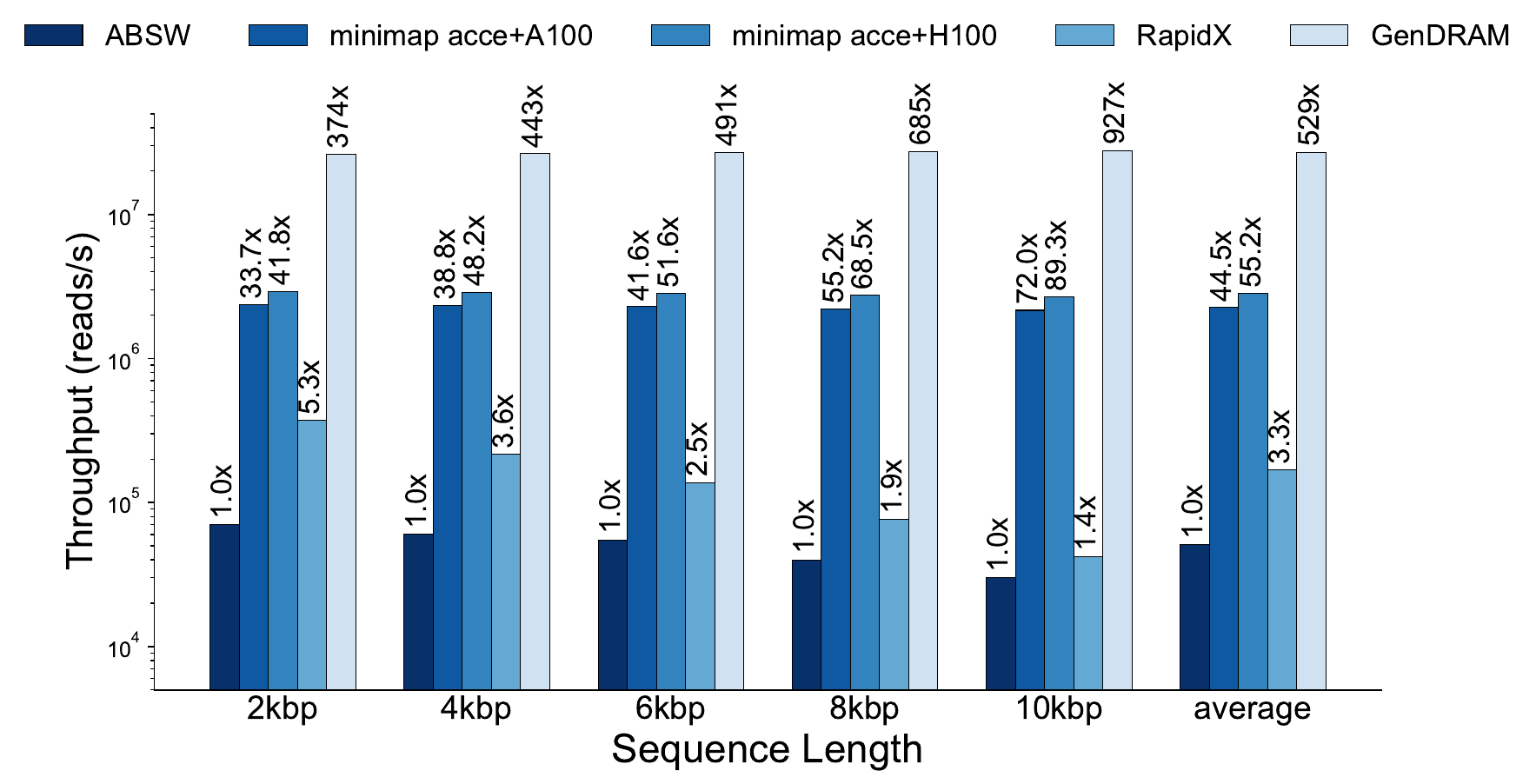}    
  \caption{Alignment throughput comparison of GenDRAM, RAPIDx~\cite{xu2023rapidx}, GASAL2~\cite{ahmed2019gasal2}, and  Minimap2 acceleration~\cite{guo2019hardwareFPGA} for long reads}
  \label{fig:genomic_perf2}
\end{figure}

For long reads (Fig.~\ref{fig:genomic_perf2}), the performance gap narrows as the workload becomes increasingly compute intensive, allowing GPUs to amortize overhead more effectively. Nevertheless, GenDRAM maintains a commanding lead with an average speedup of approximately 20$\times$ over the A100, ranging from 29$\times$ for 2kbp reads to 14$\times$ for 10kbp reads. These results validate the efficacy of GenDRAM's architecture, which leverages massive internal bandwidth and near-memory computation to effectively overcome the distinct bottlenecks inherent to both short-read (latency-bound) and long-read (compute-bound) workloads. 

Beyond GPU comparisons, GenDRAM demonstrates distinct advantages over state-of-the-art domain-specific accelerators across both workload regimes. For short reads, we compare GenDRAM against RAPIDx~\cite{xu2023rapidx} (ReRAM-based PIM) and Aligner-D~\cite{zhang2023alignerd} (DIMM-based PIM). As shown in Fig.~\ref{fig:genomic_perf1}, GenDRAM achieves approximately 15$\times$ higher throughput than RAPIDx and over 50$\times$ higher than Aligner-D. The performance gap with RAPIDx stems primarily from system-level integration: while RAPIDx optimizes the alignment kernel, its ReRAM capacity constraints force the memory-intensive seeding stage to be offloaded to the host, creating a communication bottleneck. GenDRAM's high-capacity 3D DRAM enables the entire pipeline to execute on-chip, eliminating this overhead. Similarly, Aligner-D is constrained by the standard DIMM interface bandwidth, whereas GenDRAM exploits the massive internal bandwidth of hybrid bonding.

For long reads, we benchmark against the ASIC-based ABSW~\cite{liao2018adaptivelyABSW} and GenASM~\cite{cali2020genasm}. GenDRAM outperforms ABSW by approximately 45$\times$ (Fig.~\ref{fig:genomic_perf2}). This advantage is driven by memory capacity. ABSW relies on limited on-chip SRAM, necessitating complex tiling and frequent off-chip data spills when processing long sequences. In contrast, GenDRAM's 3D DRAM stack provides ample capacity to store large wavefronts and traceback tables locally, maintaining high throughput even as sequence length increases.

\begin{figure}[htbp]
    \centering
    \includegraphics[width=1.0\columnwidth]{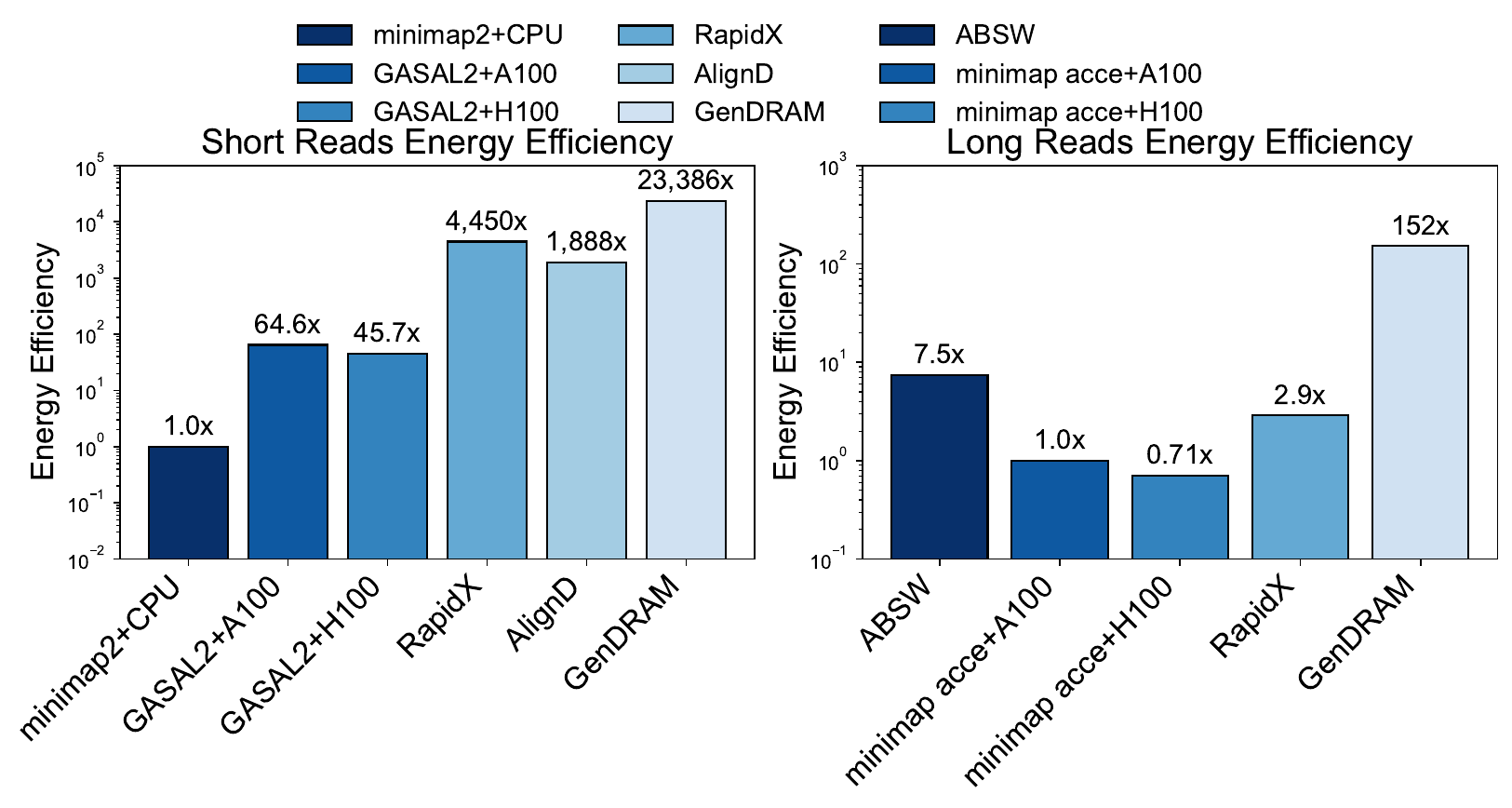}
    \caption{Energy efficiency of short and long reads}
    \label{fig:genomic_EE}
\end{figure}
\textbf{GenDRAM sequence alignment energy efficiency:}
Fig. \ref{fig:genomic_EE} presents the energy efficiency of our GenDRAM architecture compared to various CPU, GPU, and PIM-based baselines for both short-read and long-read genomic alignment.  For short-read alignment (left), results are normalized to the \texttt{minimap + CPU} baseline (1.0$\times$). Our proposed GenDRAM architecture demonstrates superior energy efficiency, achieving 23386 $\times$ that of the CPU baseline. This performance significantly outperforms other PIM-based accelerators like \texttt{RapidX} (68.9$\times$) and \texttt{AlignerD} (29.2$\times$). In contrast, the \texttt{GASAL2+H100} configuration (0.71$\times$) and the CPU-based \texttt{minimap2} (0.02x) show substantially lower energy efficiency.

A similar trend is observed for long-read alignment (right), where results are normalized to the \texttt{minimap acce+A100} baseline (1.0$\times$). GenDRAM achieves a 152$\times$ improvement in energy efficiency. This result surpasses other specialized hardware accelerators, including \texttt{ABSW} (7.5$\times$) ~\cite{liao2018adaptivelyABSW} and \texttt{RapidX} (2.9$\times$), as well as the \texttt{minimap acce+H100} (0.71$\times$).
This superior efficiency is primarily driven by the minimization of system-level data movement. Unlike the GPU baseline, which incurs substantial energy overhead transferring data between HBM and compute units, GenDRAM leverages in-situ computation to eliminate off-chip memory traffic. Furthermore, compared to domain-specific accelerators like RAPIDx and ABSW, GenDRAM avoids the energy-intensive host-device communication bottlenecks caused by on-chip capacity constraints. By integrating the massive capacity of Monolithic 3D DRAM, GenDRAM retains the entire execution context on-chip, thereby removing the costly I/O overhead that limits prior designs.

\subsection{GenDRAM Power, area and thermal Considerations}
\label{ssec:power_area_logic_die}
\begin{figure}[htbp]
    \centering
    \includegraphics[width=1.0\columnwidth]{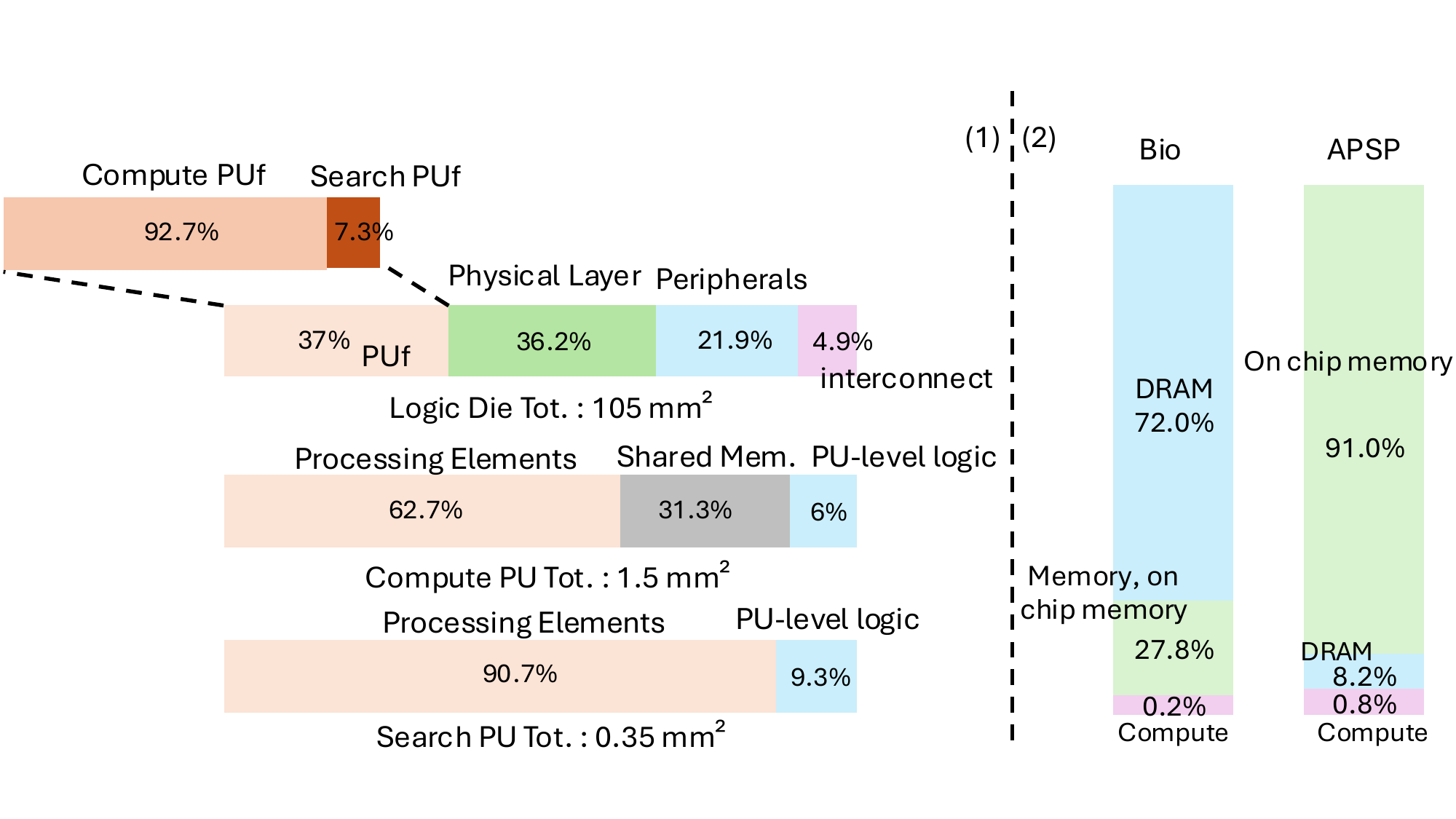} 
    \caption{(1) Area breakdown of logic die processor; (2) Power breakdown of GenDRAM logic die at peak performance}
    \label{fig:breakdown}
\end{figure}

The GenDRAM logic die, implemented on the 7nm process, integrates a heterogeneous array of 32 PUs (8 search and 24 compute PUs) operating at 1 GHz, with detailed specifications listed in Table \ref{tab:energy}. The physical implementation results are detailed in Fig.~\ref{fig:breakdown}, providing a comprehensive breakdown of both area and power consumption. 

\textbf{Area brakedown} is shown in Fig.~\ref{fig:breakdown}(1). The silicon footprint is dominated by the \textbf{light orange region} representing the Compute PUs, which occupy 92.7\% of the total processor area. This massive allocation is driven by the need for substantial on-chip SRAM (visible as the \textbf{grey blocks} labeled shared mem. within the PU breakdown) to cache data blocks during compute-bound tasks. In contrast, the Search PUs, shown in \textbf{dark orange}, utilize minimal buffering (8 KB) and occupy a negligible fraction of the area. The remaining logic die area comprises standard interfaces, dominated by the \textbf{green} Physical Layer (PHY) block. The PHY, occupying 36.2\% of the die, contains the essential transceiver circuitry required to drive the high-density hybrid bonding interconnects that link the logic die to the memory stack. Together with the \textbf{light blue} peripheral blocks, these interfaces consume approximately 58\% of the total die, reflecting the high silicon cost of sustaining massive vertical bandwidth.

\textbf{Power breakdown} is shown in Fig.~\ref{fig:breakdown}(2).  For the bioinformatics workload (left bar), the vast \textbf{light blue region} reveals that \textbf{72.0\%} of power is consumed by DRAM access, highlighting the cost of streaming massive genomic databases. Conversely, the APSP workload (right bar) is dominated by the \textbf{light green region} (\textbf{91.0\%}), which represents on-chip memory power. This distinct shift confirms that APSP effectively utilizes the local SRAM to minimize expensive DRAM access. Notably, in both workloads, the \textbf{pink regions} representing active computation account for less than 1\% of total power, reinforcing that data movement—whether off-chip (sequence alignment) or on-chip (APSP)—is the primary energy bottleneck.

\begin{table}[h]
\centering
\caption{GenDRAM Logic Die Processor Specification}
\label{tab:energy}
\footnotesize
\begin{tabular}{|c|c|c|c|} 
\hline
\multicolumn{4}{|c|}{\textbf{Processing Element (PE)}} \\
\hline
Core Logic & \makecell[c]{Adder \\  Comparator} & Tiering Table &\makecell[c]{8 x 32bit \\ Registers} \\
\hline
\makecell[c]{Compute PE \\Buffer} & \makecell[c]{32KB \\SRAM} & \makecell[c]{Search PE \\Buffer} & \makecell[c]{8 KB \\SRAM} \\
\hline
\multicolumn{4}{|c|}{\textbf{Processing Unit (PU)}} \\
\hline
\#PEs & 16 & \makecell[c]{Ring Router \\Link} & 128 GB/s \\
\hline
Shared Memory & 256 KB SRAM & \multicolumn{2}{|c|}{} \\
\hline
\multicolumn{4}{|c|}{\textbf{GenDRAM Processor}} \\
\hline
Basic & \multicolumn{3}{|c|}{\makecell[l]{7 nm process; 0.7 V supply; 105 mm$^2$ die area;\\ Int32/Int5 format}} \\
\hline
SRAM Capacity & 19.5 MB & Peak Power & 31.22W \\
\hline
\makecell[c]{Aggregated \\On-chip \\ Ring Bandwidth} & 4.096 TB/s & \makecell[c]{Aggregated \\M3D DRAM\\ Bandwidth} & \makecell[c]{19.01-\\34.34 TB/s} \\
\hline
\end{tabular}
\end{table}

In terms of power and performance, GenDRAM demonstrates high efficiency, with average power consumption of 31.2 W for the bioinformatics pipeline and 10.15 W for APSP, remaining well within the thermal budget of a PIM system ~\cite{woon2025thermal, kim2016neurocube}. Power dissipation is workload-dependent: the bioinformatics workload is dominated by massive parallel DRAM accesses during seeding, whereas APSP is governed by energy-efficient SRAM accesses (0.007 nJ/access for local memory). Efficiency is further enhanced by multiplier-less PEs and a programmable tiering table that dynamically manages access latencies across DRAM layers. A high-speed on-chip ring network providing 128 GB/s per link facilitates the deep pipelining for bioinformatics and broadcasting for APSP, validating the architecture's significant performance-per-watt advantage ~\cite{pan2025stratum, xu2023rapidx}.

\subsection{Sensitivity and scalability analysis}
In this section, we expand beyond baseline performance metrics to conduct a comprehensive design space exploration, validating the robustness of GenDRAM's architectural decisions against physical constraints. We focus on quantifying the impact of three critical hardware trade-offs: the non-uniform latency of the 3D stack, the partitioning of heterogeneous processing resources, and the sizing of on-chip local memory. Our analysis first highlights the efficacy of the proposed 3D-aware mapping strategy, demonstrating that prioritizing latency-critical data for bottom tiers is imperative to hide the access penalty of upper DRAM layers, thereby recovering performance levels comparable to idealized uniform-latency memory. Furthermore, regarding resource partitioning, our sweep of Processing Unit (PU) configurations validates that the selected ratio of 8 Search PUs to 24 Compute PUs represents the sweet point. This specific configuration perfectly balances the throughput of the memory-bound Seeding phase and the compute-bound Alignment phase, minimizing pipeline stalls to maximize end-to-end utilization. 

\subsubsection{\textbf{Tiering and latency heterogeneity in M3D DRAM}}
\label{ssec:tiering_in_3d_dram}

A fundamental physical characteristic of the aggressively scaled, 1024-layer M3D DRAM is the significant variation in access latency across its vertical layers. As detailed in the Stratum paper and illustrated in our background section, this latency heterogeneity stems from the wordline (WL) staircase structure required for routing ~\cite{pan2025stratum}. WLs routed to layers farther from the logic die must traverse a longer path with a greater number of metal segments, leading to a linear increase in resistance and capacitance (RC) delay ~\cite{pan2025stratum}.

Our simulation model, configured with the parameters from Table \ref{tab:3ddram_params}, faithfully captures this effect. The row-to-column delay $(t_{RCD})$ ranges from just 2.29 ns for the fastest tier to 22.88 ns for the slowest tier. When considering the full row access latency $(t_{RC} = t_{RP} + t_{RAS})$), this translates to a significant performance differential. The fastest tier can complete a full row access in approximately 34.56 ns, while the slowest tier requires 55.15 ns~\cite{pan2025stratum}. This confirms that the fastest tier provides nearly 1.6x faster memory access than the slowest tier, a key characteristic that our data mapping strategy is designed to exploit ~\cite{pan2025stratum}. Rather than being a limitation, we leverage this predictable latency variation by mapping latency-sensitive data structures, such as the PTR and CAL tables for Seeding, to the faster, lower tiers, thereby optimizing the end-to-end pipeline performance.

\begin{figure}[htbp]
  \centering
  \includegraphics[width=1.0\columnwidth]{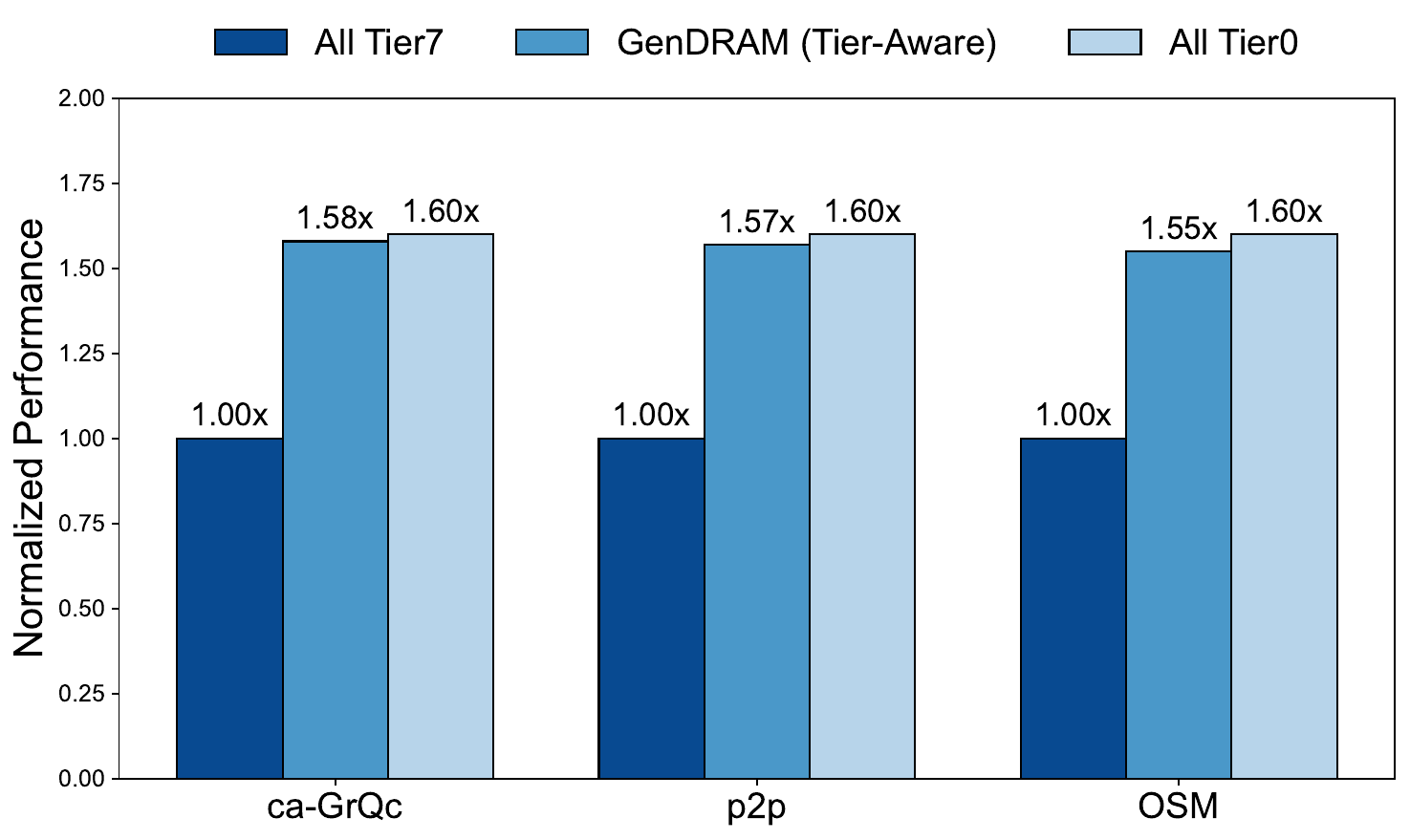}   
  \caption{Different mapping strategy speedup}
  \label{fig:diffmapping}
\end{figure}

To quantify the impact of this 3D-aware mapping strategy, we performed a sensitivity analysis comparing GenDRAM against two extreme baselines, as shown in Fig.~\ref{fig:diffmapping}. The \textbf{Uniform Worst-Case (All Tier 7)} baseline assumes a naive mapping where all memory accesses incur the maximum latency of the top-most layer (55.15 ns), representing a standard architecture unaware of vertical variation. Conversely, the \textbf{Uniform Best-Case (All Tier 0)} represents a theoretical upper bound where the entire memory stack operates at the speed of the bottom-most layer (34.56 ns). Results demonstrate that the naive "All Tier 7" mapping degrades performance significantly, serving as the 1.0$\times$ baseline. GenDRAM's tier-aware approach achieves speedups of approximately \textbf{1.58$\times$}, recovering nearly \textbf{98\%} of the performance of the idealistic "All Tier 0" case (1.60$\times$). This confirms that our software-hardware co-design effectively hides the latency penalties of the upper layers. By pinning the latency-critical Pointer (PTR) and Candidate Location (CAL) tables to Tier 0, GenDRAM ensures that the frequent random accesses in the Seeding phase are serviced with minimal delay, while the bandwidth-bound streaming data (which is less sensitive to latency) utilizes the upper capacity.

\subsubsection{\textbf{Resource Partitioning: Search vs. Compute PUs}}
\label{ssec:pu_partitioning}

\begin{figure}[htbp]
  \centering
  \includegraphics[width=1.0\columnwidth]{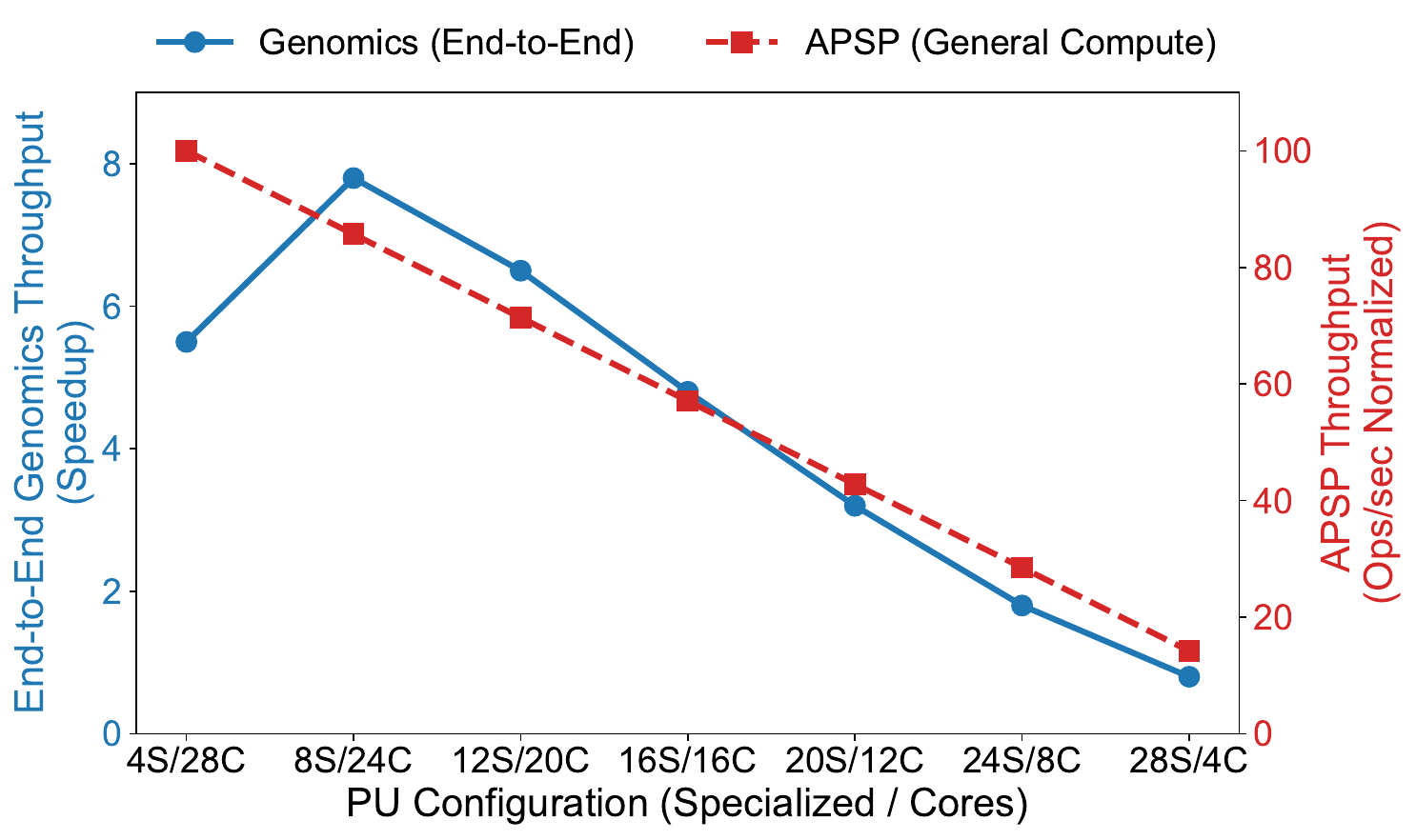}   
  \caption{Different configuration strategy speedup}
  \label{fig:diffSC}
\end{figure}

Given the fixed silicon area of 105 mm$^2$, a critical architectural trade-off involves partitioning resources between the heterogeneous Search PUs (for memory-bound Seeding) and Compute PUs (for compute-bound Alignment/APSP). We explored the design space by sweeping the ratio of a fixed 32-PU cluster. 
A search and heavy configuration (e.g., 16 Search / 16 Compute) accelerates the seeding phase but creates a severe bottleneck in the subsequent alignment phase, leaving search PUs idle once the buffer fills and degrading end-to-end throughput. 
Conversely, a compute and heavy configuration (e.g., 4 Search / 28 Compute) maximizes theoretical peak performance for APSP but starves the pipeline during genomic workloads, as the few Search PUs cannot produce candidate seeds fast enough to keep the massive compute array busy. 
Our analysis, shown on Fig. ~\ref{fig:diffSC}, identifies the 8 Search PUs to 24 Compute PUs ratio as the sweet point. This 1:3 ratio aligns with the workload characteristics where Seeding accounts for roughly 25-30\% of the pipeline latency in our accelerated model, ensuring a balanced producer-consumer flow that maximizes overall system utilization without wasting silicon area on idle units.

\subsubsection{\textbf{System-level pipeline configurations analysis}}
\begin{figure}[htbp]
  \centering
  \includegraphics[width=1.0\columnwidth]{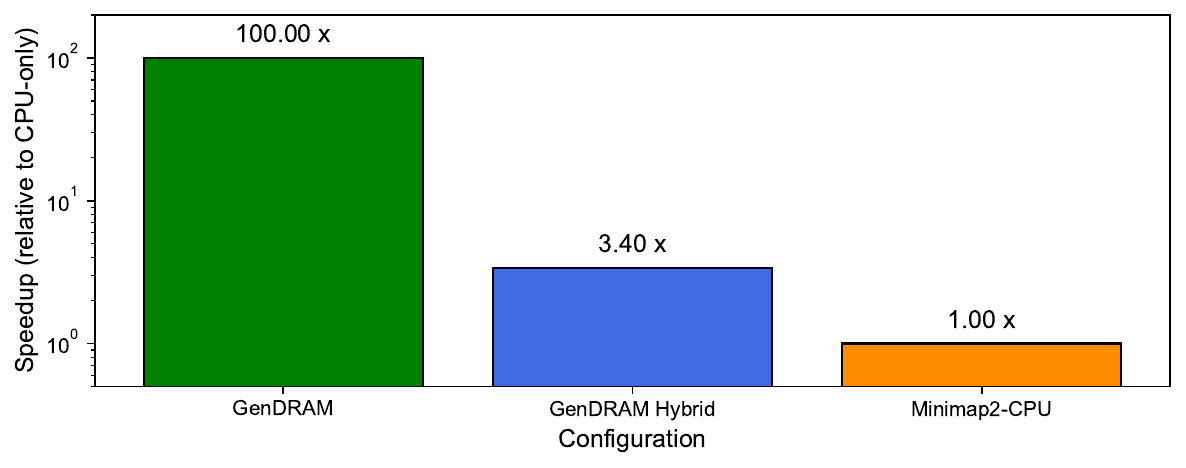}   
  \caption{Sequence alignment configurations performance}
  \label{fig:pipeline_comparison}
\end{figure}

To quantify the system-level bottlenecks inherent in traditional and hybrid computing models, we evaluated three distinct pipeline configurations for the genomics workflow. The end-to-end performance results, shown in Fig.~\ref{fig:pipeline_comparison},  compare a Minimap2-CPU ~\cite{li2018minimap2}, a hybrid model where seeding is performed on the host CPU and alignment is offloaded to GenDRAM, and the fully integrated GenDRAM design. 

The Minimap2-CPU baseline establishes the reference performance (1.00$\times$). Profiling of this workload reveals that execution time is split between the memory-intensive Seeding stage ($\sim$30\%) and the compute-intensive alignment stage ($\sim$70\%)~\cite{li2018minimap2, ahmed2019gasal2}. In the GenDRAM Hybrid model, we offload only the alignment stage to the accelerator while retaining seeding on the host CPU. Despite GenDRAM's massive compute capability, this configuration achieves only a 3.40$\times$ speedup. The unaccelerated seeding stage on the host becomes the dominant bottleneck. Specifically, while GenDRAM processes the alignment rapidly, it remains idle for approximately 90\% of the total execution cycles, waiting for the slower CPU to produce seeds. This highlights the inefficiency of moving data across the PCIe bus and splitting the pipeline.

The fully integrated GenDRAM approach (green bar) overcomes this bottleneck by executing the entire pipeline on-chip. By accelerating both seeding (via Search PUs) and alignment (via Compute PUs) within the PIM architecture, we eliminate the host serialization point. This results in a 100$\times$ speedup over the CPU baseline and a 29$\times$ improvement over the Hybrid model. This quantitative gap validates our core thesis: true acceleration for complex genomic pipelines requires eliminating host-accelerator communication and processing all stages near memory.

\subsubsection{\textbf{GenDRAM PU and PE performance tradeoffs}}
\label{PUPE_tradeoff}

To contextualize the scaling analysis, we briefly recapitulate the GenDRAM logic die architecture. The processor is organized as a cluster of PUs, serving as the top-level building blocks. Each PU couples a dedicated local SRAM (512 KB) with a systolic array of PEs, the fundamental arithmetic units responsible for executing the massive parallel operations. 

Building on the resource partitioning analysis in Section~\ref{ssec:pu_partitioning}, which established the 1:3 ratio (8 Search PUs to 24 Compute PUs) as the best configuration for balancing the seeding and alignment pipeline stages, we now evaluate the scalability of the total system size while maintaining this optimal ratio.

\begin{figure}[htbp]
  \centering
  \includegraphics[width=1.0\columnwidth]{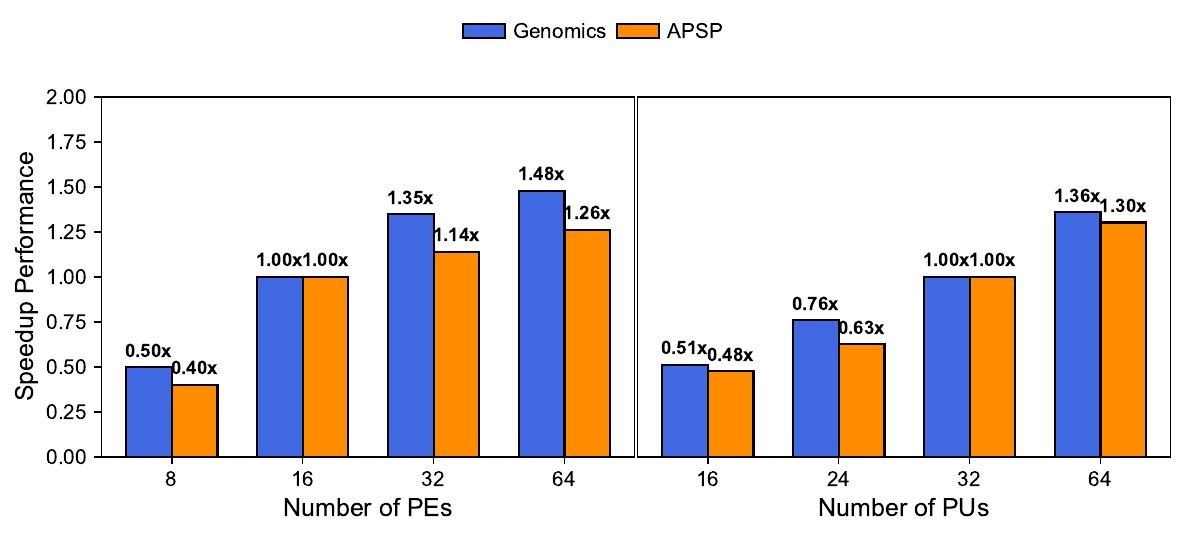}    
  \caption{Impact of different PU and PE configurations on execution time}
  \label{fig:pu_scaling_perf}
\end{figure}

We performed a design space exploration simulating configurations with 16, 24, 32, and 64 PUs across genomics and APSP workloads. As illustrated in Fig.~\ref{fig:pu_scaling_perf}, the results reveal a clear trade-off between computational parallelism and memory system limitations. For the compute-bound genomics pipeline, performance scales well, with relative performance nearly doubling from 16 to 32 PUs (0.51$\times$ to 1.00$\times$) and reaching 1.36$\times$ with 64 PUs. Similarly, the APSP workload demonstrates significant performance improvement, more than doubling from 0.48$\times$ with 16 PUs to 1.00$\times$ with 32 PUs, identifying 32 PUs as the optimal design point.

The selection of 32 PUs is architecturally driven to match the parallelism of the underlying M3D DRAM, which is organized into 32 independently accessible bank-groups. A 16-PU configuration would underutilize the memory system, leaving half of the available memory bandwidth idle. Conversely, scaling to 64 PUs leads to diminishing returns due to memory contention; with 64 PUs competing for only 32 parallel memory pathways, memory access becomes the system bottleneck, causing processor stalls. Therefore, 32 PUs represents the architectural sweet spot at the chip level.

In addition to the chip-level PU count, the number of PEs within each PU is a critical microarchitectural design choice. We selected 16 PEs per PU to co-design the internal compute fabric with the data access granularity of the DRAM system. A single DRAM row activation provides a wide data block of 8192 bits, which corresponds directly to the vector size of our key algorithms, such as a 256-element row in an APSP tile (256 elements $\times$ 32 bits). A cluster of 16 PEs is perfectly sized to process this 8192-bit data vector in parallel, with each PE handling a 512-bit slice.

While hypothetically increasing the PE count to 32 or 64 might enhance raw computational throughput, such a design encounters severe bottlenecks in physical area and intra-PU bandwidth. Scaling the PE count causes unsustainable growth in PU area, driven by the tightly coupled 32KB local SRAMs; doubling PEs to 32 would nearly double the total PU area, making it impossible to integrate the required 24 Compute PUs within the logic die's area budget. Furthermore, an intra-PU memory bandwidth bottleneck would emerge, as the single 256KB shared memory cannot service twice the number of compute units without saturation, leading to widespread stalls that negate computational benefits.

Finally, a significant increase in PE density creates a larger power and thermal load. Doubling the active PEs and their associated SRAMs within the confined area of a PU leads to a projected $\sim$2$\times$ spike in local power density. Given that the logic die is situated directly beneath the heat-sensitive M3D DRAM stack, such thermal hotspots pose a significant risk to the reliability and data integrity of the memory above. Thus, 16 PEs per PU is the optimal choice, representing a carefully balanced co-design of compute parallelism, memory access, area, and power for an efficient and feasible microarchitecture.


We further analyzed the impact of internal parallelism by sweeping the number of PEs per Compute PU, as illustrated in the left panel of Fig.~\ref{fig:pu_scaling_perf}. The results reveal a distinct saturation point in performance scaling.
Increasing the density from 8 to 16 PEs yields a near-linear speedup, boosting relative performance from 0.50$\times$ to 1.00$\times$. However, scaling beyond 16 PEs results in severe diminishing returns due to local SRAM bandwidth contention and data dependency limits. As shown in the figure, doubling the resources to 32 PEs provides only a marginal performance gain—improving Genomics throughput by just 35\% and APSP by merely 14\%, despite incurring a 100\% increase in logic area and power consumption.

This trade-off is critical for thermal feasibility. Moving from 16 to 32 PEs would lead to a projected $\sim$2$\times$ spike in local power density within the confined PU footprint. Given that the logic die is vertically stacked directly beneath the heat-sensitive 3D DRAM, creating such concentrated thermal hotspots poses a significant risk to memory reliability. Consequently, 16 PEs per PU represents the Pareto-optimal design point, capturing the knee of the performance curve while maintaining a power density compatible with the thermal constraints of monolithic 3D integration.

\subsection{GenDRAM overhead analysis}
While GenDRAM delivers transformative performance and efficiency, these gains come with specific architectural and physical costs. We analyze these overheads in terms of silicon area, thermal management, and manufacturing complexity.

\subsubsection{\textbf{Silicon area overhead}}
The integration of the logic die at the base of the memory stack introduces an additional silicon footprint of 105 mm$^2$. As detailed in Fig.~\ref{fig:breakdown}, a significant portion of this area is dedicated to infrastructure rather than computation. Specifically, the PHY and peripheral interfaces occupy $\sim$58\% of the die area. This is required to drive the high-density hybrid bonds and sustain the massive vertical bandwidth of the M3D stack. However, when viewed at the system level, this overhead is minimal. Compared to the massive 826 mm$^2$ die size of the NVIDIA A100 GPU, GenDRAM's logic layer requires only $\sim$12.7\% of the silicon area while delivering superior performance for targeted workloads, representing a highly area-efficient design trade-off.

\subsubsection{\textbf{Thermal constraints}}
The primary operational cost of the monolithic 3D architecture is the increased thermal density. Placing active logic directly beneath DRAM layers creates a thermal coupling effect not present in standard 2D DRAM. Our simulations indicate that the logic layer generates a power density peak of roughly 0.3 W/mm$^2$ (at 31.21 W total power). While this necessitates careful thermal management to prevent data retention errors in the DRAM refresh rows, our analysis in Section~\ref{ssec:pe_scaling} confirms that limiting the design to 16 PEs per PU keeps this overhead well within the safe operating limits of standard passive cooling solutions ($<$15 W/stack thermal budget for typical use cases, extensible with active cooling).

\subsubsection{\textbf{Manufacturing complexity}}
Finally, GenDRAM relies on advanced monolithic 3D fabrication processes utilizing sequential tier integration and hybrid bonding. This incurs a higher manufacturing cost and yield risk compared to traditional TSV based HBM or standard DDR4 DIMMs. However, this manufacturing overhead is justified by the key benefits: M3D provides orders-of-magnitude higher vertical interconnect density ($\sim$100 million connections/mm$^2$) and lower access energy. Furthermore, these costs are expected to fall as the manufacturing process matures.


\section{Conclusion}
In this work, we presented GenDRAM, a PIM accelerator designed to address the system-level bottlenecks inherent in data-intensive dynamic programming applications. By co-designing a heterogeneous architecture of specialized Search and Compute processing units with the unique properties of M3D DRAM, we demonstrated the feasibility of integrating entire complex workflows, such as genomic sequence alignment and graph-based APSP onto a single chip. This approach eliminates the critical host-accelerator communication overhead and fully exploits the massive internal bandwidth and tiered-latency characteristics of 3D-stacked memory.  Compared to the state-of-the-art NVIDIA A100 GPU, GenDRAM delivers speedups of 68$\times$ for APSP and 22$\times$ for end-to-end genomic alignment, while improving energy efficiency by 152$\times$. Crucially, GenDRAM also significantly outperforms specialized domain-specific accelerators. By overcoming the write-latency and capacity limitations of ReRAM-based designs, GenDRAM achieves 20$\times$ higher energy efficiency than RapidGraph for graph analytics and provides a 15$\times$ to 45$\times$ throughput advantage over RAPIDx and ABSW for genomic workloads. Ultimately, GenDRAM shows that a deeply co-designed M3D PIM architecture is a powerful and practical solution for the escalating challenges of real-time, large-scale data analytics.

{
    \small
    \bibliographystyle{ieeetr}
    \bibliography{reference}
}

\vskip -3.0\baselineskip plus -1fil

\begin{IEEEbiography}[{\includegraphics[width=0.9in,clip,keepaspectratio]{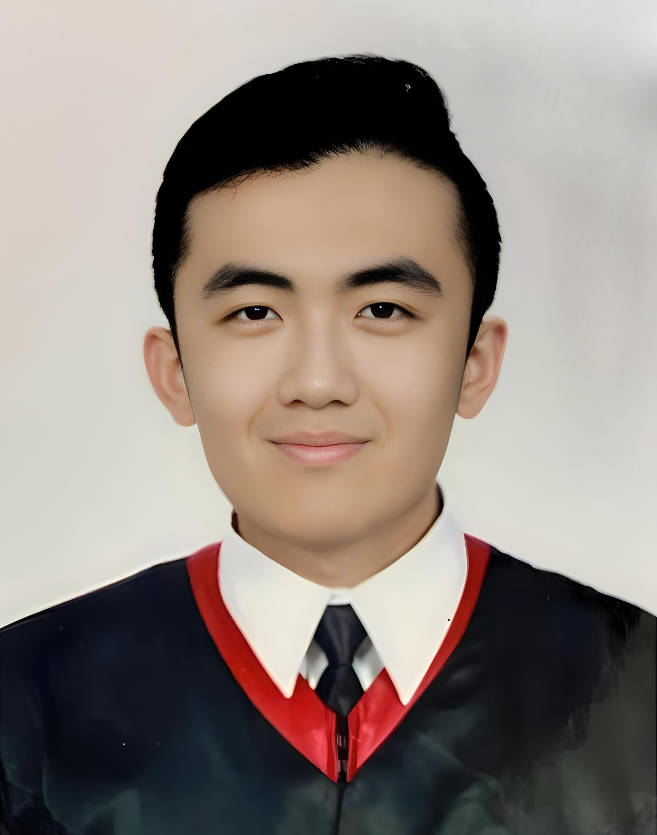}}]{Tsung-Han Lu}
received M.s in Power Mechanical Engineering from National Tsing Hua University, HsinChu, Taiwan, in 2021. He is currently a Second-year Ph.D. student in Electrical and Computer Engineering at the University of California San Diego, La Jolla, CA, USA. 
His research interests include in-memory and in-storage architecture for deep learning, bioinformatics. 
\end{IEEEbiography}

\vskip -3.0\baselineskip plus -1fil
\begin{IEEEbiography}[{\includegraphics[width=0.9in,clip,keepaspectratio]{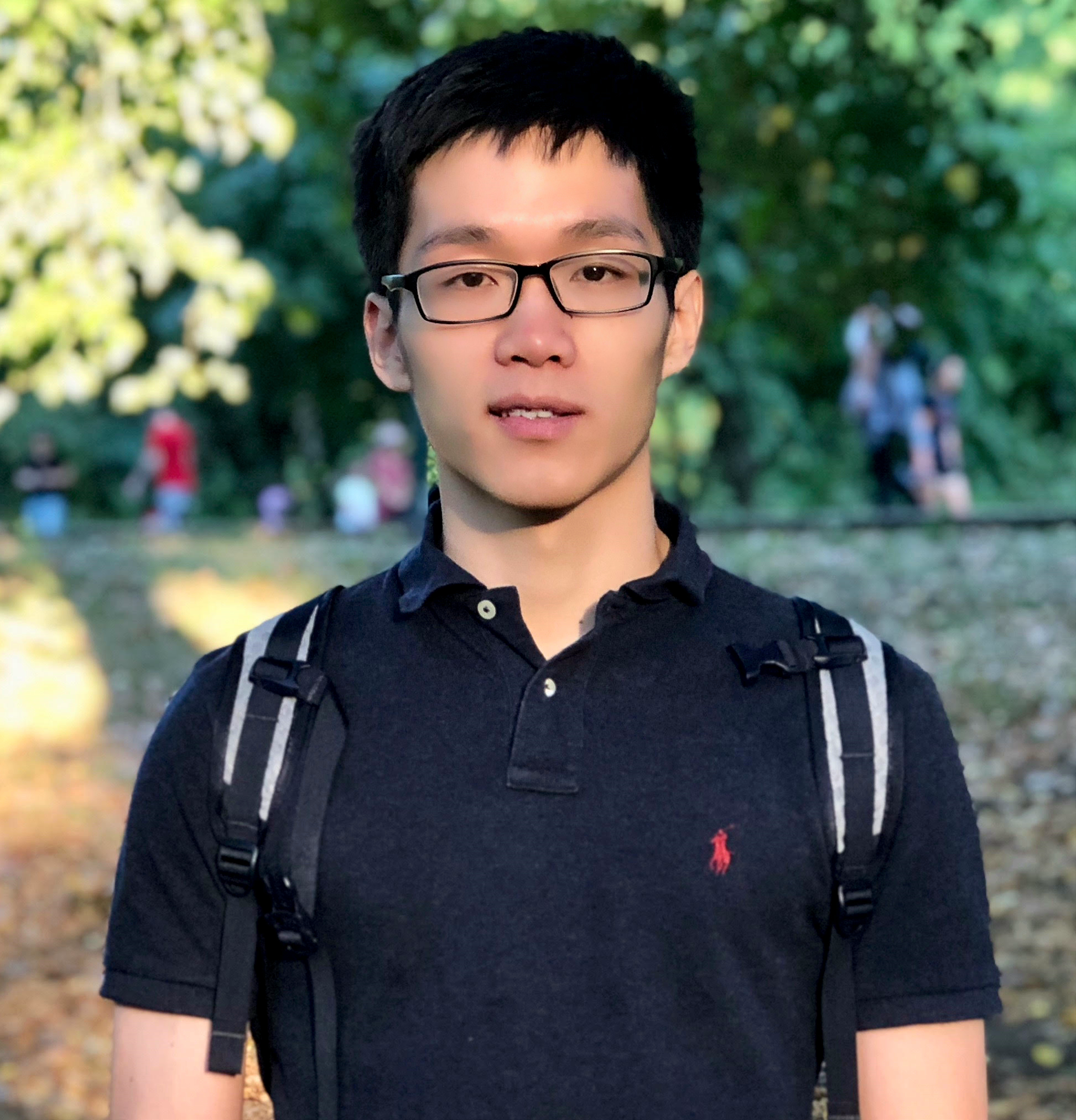}}]{Weihong Xu}
received the B.E. degree in information engineering and M.E. in information and communication engineering from Southeast University, Nanjing, China, in 2017 and 2020. Received PhD degree in Computer Engineering at the University of California San Diego, La Jolla, CA, USA. He is currently a Postdoc in École polytechnique fédérale de Lausanne. His research interests include in-memory and in-storage architecture for deep learning, bioinformatics, and hyperdimensional computing. 
\end{IEEEbiography}

\vskip -2.5\baselineskip plus -1fil

\begin{IEEEbiography}[{\includegraphics[width=0.9in,height=1.2in,clip,keepaspectratio]{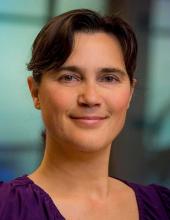}}]{Tajana Rosing}
received her Ph.D. degree from Stanford University, Stanford, CA, USA, in 2001. She is a Professor, a Holder of the Fratamico Endowed Chair, and the Director of System Energy Efficiency Laboratory, University of California at San Diego, La Jolla, CA, USA. From 1998 to 2005, she was a full-time Research Scientist with HP Labs, Palo Alto, CA, USA, while also leading research efforts with Stanford University, Stanford, CA, USA. She was a Senior Design Engineer with Altera Corporation, San Jose, CA, USA. She is leading a number of projects, including efforts funded by DARPA/SRC JUMP 2.0 PRISM program with focus on design of accelerators for analysis of big data, DARPA and NSF funded projects on hyperdimensional computing, and SRC funded project on IoT system reliability and maintainability. Her current research interests include energy-efficient computing, cyber–physical, and distributed systems.
\end{IEEEbiography}

\end{document}